\documentstyle[aps,epsf,rotate]{revtex}
\begin{document}
\title{A comprehensive numerical and analytical study \\
of two holes doped into the 2D $t-J$ model}

\author{A. L. Chernyshev\cite{perm}}
\address{Dept. of Physics, Queen's University, Kingston, Ontario, 
Canada K7L 3N6}
\author{P. W. Leung}
\address{Physics Dept., Hong Kong University of Science and Technology,
Clear Water Bay, Hong Kong}
\author{R. J. Gooding}
\address{Dept. of Physics, Queen's University, Kingston, Ontario, 
Canada K7L 3N6}

\date{\today} 
\maketitle

\begin{abstract}
We report on a detailed examination of numerical results and
analytical calculations devoted to a study of two holes doped into
a two-dimensional, square lattice described by the $t-J$ model.
Our exact diagonalization numerical results represent the first solution
of the exact ground state of 2 holes in a 32-site lattice. Using
this wave function, we have calculated several important correlation
functions, notably the electron momentum distribution function
and the hole-hole spatial correlation function. Further, by studying
similar quantities on smaller lattices, we have managed to perform
a finite-size scaling analysis. We have augmented this work by
endeavouring to compare these results to the predictions of
analytical work for two holes moving in an infinite lattice. This analysis 
relies on the canonical transformation approach formulated recently 
for the $t-J$ model. From this comparison we find excellent correspondence 
between our numerical data and our analytical calculations. We 
believe that this agreement is an important step helping to justify 
the quasiparticle Hamiltonian, and in particular, the quasiparticle 
interactions, that result from the canonical transformation approach. Also, 
the analytical work allows us to critique the finite-size scaling ansatzes 
used in our analysis of the numerical data. 
One important feature that we can infer from 
this successful comparison involves the role of higher harmonics in the 
two-particle, $d$-wave symmetry bound state --- the 
conventional $(\cos(k_x) - \cos(k_y))$ term is only one of many 
important contributions to the $d$-wave symmetry pair wave function.

\end{abstract}

\section{Introduction}

The behaviour of mobile holes in an antiferromagnetic (AF) spin background
has been a subject of intensive study, in part because of its 
possible connection to high temperature superconductivity. The ubiquitous
structural components of such materials are the ${\rm CuO_2}$ planes,
and the description of carriers in these planes is the theoretical focus
of this paper. We consider the so-called $t - J$ model \cite {pwa87,elbio94}, 
for which the holes correspond to the Zhang-Rice singlets \cite{ZR},
mobile vacancies 
created by doping a single $\rm CuO_2$ plane.
A microscopic representation of this model is
\begin{equation}
{\cal H}_{t-J} = -t\sum_{\langle ij\rangle\sigma}(\tilde{c}^\dagger_{i\sigma}
\tilde{c}_{j\sigma}+{\rm H.c.})+J\sum_{\langle ij\rangle} 
({\bf S}_i\cdot {\bf S}_j
-\frac{1}{4}n_in_j),
\label{hamiltonian}
\end{equation}
where $\langle ij\rangle$ denotes nearest neighbour sites,
and $\tilde{c}^\dagger_{i\sigma}$, $\tilde{c}_{i\sigma}$ are the constrained
operators, $\tilde{c}_{i\sigma}=c_{i\sigma}
(1-c^\dagger_{i,-\sigma}c_{i,-\sigma})$. The ratio of the AF exchange constant 
to the hopping amplitude is believed to be $J/t \sim 0.3$.

Aided by recent angle-resolved photoemission experiments \cite {wells},
followed by extensive comparisons between theory and experiment \cite
{wellscomp}, 
it is now recognized that this simple,
near-neighbour hopping Hamiltonian on its own is insufficient to
fully represent the true single hole state of the real $\rm CuO_2$
plane. Hoppings between more
distant neighbours are required \cite {wellscomp,1bndNaz,1bndBel}, 
as are more
complicated three-site spin-dependent hoppings. Possibly,
the full three-band microscopic Hamiltonian is necessary \cite{3bnd}.

Despite the potential 
inadequacy of this Hamiltonian to represent completely the
$\rm CuO_2$ planes, it is still the simplest model that 
captures the important antiferromagnetic correlations of a
weakly doped antiferromagnet \cite {pwa97}.
Thus, it is crucial that the properties of this model 
when doped are elucidated. 

The Hamiltonian in Eq. (\ref{hamiltonian}) has been investigated intensively
by different analytical and numerical methods, and we believe that a 
consistent picture has emerged from these studies. 
Some results of the exact diagonalization on small
clusters, such as the energy spectrum and quasiparticle residue,
have been found in the excellent agreement with the results
predicted by advanced analytical theory \cite {lg95}. In contrast with this,
a large amount of the numerical data for the correlation functions
are not well understood or require further explanation. 
Such an explanation, when completed, could help to
define (or justify) the correct quasiparticle model for the system of
strongly interacting holes and spins at low energies. It is in this
manner that we unite our analytical and numerical work in this paper.

A common and apparently successful theory of a single hole moving
in an AF aligned background involves the so-called spin polaron
\cite {BNK,KLR,Man1,Horsch}.
According to the spin-polaron idea, the hole in its movement disturbs
the magnetic background that one can formally describe as the strong
coupling of the hole and spin degrees of freedom. This makes this problem
similar to the well known strong coupling electron-phonon one.
However, in spite of the qualitative similarity of these two
polarons, there is an essential difference between them.  If the phonon
polaron can be considered as an almost static object of the shifted
ions with the electron in the centre, the spin polaron is the
``spin-bag" with the {\it moving} hole inside.
One of the statements of the present paper is that this feature of the
spin polaron is responsible for the absence of the direct similarity
between the answers which theory provides for {\it quasiparticles} and the
numerically obtained data for {\it bare} holes. A similar conclusion,
using a different analytical approach and numerical results for
smaller clusters, was reached by Eder {\it et al} \cite {EB,EW,Ed}. 
Later, similar
remarks were made by Riera and Dagotto \cite{RD}.

In this paper we combine analytical and exact diagonalization (ED) numerical 
results of the one- and two-hole problem to provide a comprehensive
study of these  important systems. Computationally we have managed,
for the first time, to determine the two-hole ground state for two holes doped
into the 32-site, square cluster used by two of us in a previously
published numerical work \cite {lg95}. 
We find that the lowest energy state is a ${\bf P}= 0$ bound state 
with $d_{x^2-y^2}$ symmetry. 
We have characterized the ground state by
evaluating a number of important expectation values, notably
the electron momentum distribution function (EMDF), and the spatial
pair correlation function.

We have found that an effective quasiparticle Hamiltonian, originally 
proposed by Belinicher, one of us, and Shubin \cite {bel97}, may be used 
to calculate the same expectation values that were obtained numerically via 
ED. Further, these quantities are remarkably similar to those obtained via ED. 
This gives strong support to the appropriateness of this quasiparticle 
Hamiltonian. 

Previous analytical work on the low-energy physics of 
the two-hole system generally describes it in terms of moderately
interacting spin polarons. This analytical work shows that the dominant 
effective interactions between spin polarons come from the short-range
nearest-neighbour static attraction and spin-wave exchange, the latter
leading to a long-range dipolar interaction. These interactions are 
attractive for $d$-wave states and strongly repulsive for $s$-wave states.
The purpose of this paper is to use the ED results to provide
support for this description of the internal structure of the quasiparticles,
and indirectly for the above-mentioned description \cite {bel97} of their 
interactions.

We will first describe the present status of the $t-J$ model studies
in \S II.  Section III discusses in detail the numerical data available
for the ground state correlation functions. 
Then, \S IV summarizes briefly the analytical results of relevant previous
work and displays the details of the present calculations.
Section V focuses on the comparison of the analytical and numerical results,
and in \S VI we present our conclusions.

\section{Previous ${\lowercase {t}}-J$ model studies:}
\subsection{Analytical results}

Theoretical studies of the $t-J$ model have resulted in a
clear understanding of the nature of the low-energy excitations for the system
near half filling.  The charge carrier created by a hole introduced in an
AF background is described as a spin polaron, {\it viz.} as a quasiparticle 
consisting of a hole and a cloud of spin excitations. The AF 
spin-polaron concept was put forward in earlier works of 
Bulaevskii {\it et al} \cite{BNK} and Brinkman {\it et al} \cite{BR}, 
and then developed in a number of more recent papers 
\cite{KLR,Man1,Horsch,SVR,FM,Trug,GR,Eder1,DS,Sush1,Dots} 
using different techniques.

The main conclusion of these papers was that the spin polaron in an AF
background is a well defined quasiparticle with a nonzero residue and a 
specific dispersion law. The dressing of the hole leads to the narrow 
quasiparticle band with a bandwidth $\sim 2J$ for realistic $J/t$, 
band minima at ${\bf k} = \pm(\pi/2,\pm\pi/2)$, and a heavy effective
mass along magnetic Brillouin zone (MBZ) boundary.
Most of these features of the spin polaron were found to be robust under
generalizations of the $t-J$ model \cite{Zaan}, including further neighbour
and three-site hoppings, and for a wide range of $J/t$ ratio.

The single-hole problem has been treated analytically in
detail 
\cite{BNK,KLR,Man1,Horsch,BR,FM,Trug,GR,Eder1,DS,Sush1,SS2}.
Grouping these efforts, two approaches in treating this problem were
used: (i) the self-consistent Born approximation (SCBA) ({\it e.g.},
see Refs. \cite{Man1,Horsch,FM}), 
and (ii) the so-called ``string'' approach ({\it e.g.}, see Refs.
\cite{BNK,BR,Trug,Eder1,DS,Sush1}). Recently, a relationship between these two 
has been established \cite{Star}. We briefly review these methods
with an eye to understanding how well they might be able to describe
the two or multi-hole problem.

The SCBA method utilizes a property of the hole-magnon interaction, namely
the absence of the lowest order correction to the Born approximation 
series for the single-particle Green's function \cite{Man1,Horsch,FM}. 
The attractive feature of the SCBA approach is that the
essentially exact single-hole spectral functions can be evaluated quite
easily using simple numerical calculations. Recently, the detailed structure 
of the single-hole ground state and behaviour of different correlators has 
been studied using SCBA \cite{RH}.  Unfortunately, already in two- or
many-hole problems much more involved numerical and analytical 
efforts are required \cite{Plakida}.

The string approach is based on the idea that in
an AF background a hole will be confined by an effective potential
created by overturned spins (``strings''). Formally, the real-space
variational ansatz for the polaron's wave function 
with the strings of different length is considered to reproduce 
this tendency. In spite of the considerable success of the string
approach for the single-hole problem \cite{Eder1}, there are some
problems which make the use of it as a candidate for a quasiparticle 
theory for the $t-J$ model questionable. 
First of all, the string method uses the real-space approach
which does not treat properly the long-range dynamics of the system. 
The next problem is that the method starts from the Ising background
and includes fluctuations on a perturbative basis,
whereas the fluctuations are strong in 2D and must be ``build
in'' to the real ground state of the spin system. 
The third problem concerns the necessity of the normal anticommutation
relations of the quasiparticle operators. If hole is a fermion  
a unitary transformation, which diagonalizes the Hamiltonian
and dresses the hole by the spin excitations, would not change its
statistics. Within the variational ``string'' and other approaches,
one works with the wave functions and usually identifies the wave
function of the quasiparticle with the {\it operator} of the
quasiparticle. This leads to the absence of the commutation relations
for these operators and to troubles with the proper normalization and
orthogonality of the states already for the two-hole problem 
\cite{RD,EderOhta}. Because of that one cannot correctly derive the
effective polaron-polaron interaction term using a
single-hole wave function.

These difficulties notwithstanding, several attempts to address the two-hole 
ground state have been made. Work based on
the string approach have led to some qualitative understanding of
the problem \cite{Ed,Trug,SS1}. 

The investigation of the interactions between
quasiparticles in the $t-J$ model is a subject of
prime interest in the context of magnetic pairing mechanism. 
However, studies of this problem show much less convergence than the 
single-hole problem. Other work involves the formulation of an effective
model for spin polarons propagating in the AF background
\cite{SS2,SK,SWZ,FKS,NazD}. From the RPA treatment of the Hubbard model
in the strong-coupling limit, the model of ``spin-bags''
interacting via longitudinal magnetization fluctuations has been
proposed \cite{SWZ}. A phenomenological model for the vacancies coupled
by the long-range dipolar twist of the spin background has been also
worked out \cite{SS2,SS3} using a semi-classical
hydrodynamic approach. Inspired by the numerical evidence of the
hole-hole $d$-wave bound state and the Van Hove singularity in the
single-hole spectrum, taking the simplest phenomenological form of the
interaction, an AF Van Hove model has been put forward \cite{NazD}.
Using an ansatz for the quasiparticle wave function \cite{Sush1}, the
``contact'' hole-hole and the residual hole-magnon interactions have
been obtained \cite{SD,Svert}, and then the effective Hamiltonian for
the polarons and long-range spin-waves has been presented \cite{SK}. 

Even though most of these theories were formulated on a
phenomenological and semi-phenomenological basis, they provided 
two key interactions leading to pairing in the $t-J$ model. 
One of them is the effective hole-hole static attraction 
due to minimization of the number of broken bonds found from placing 
two holes at nearest-neighbour sites
(sometimes referred to as the ``sharing common link effect''). 
The other is due to
spin-wave exchange, and leads to a dipolar-type interaction between holes
\cite{SS2,SK,Fren}.

Quite recently, a new approach to the derivation of a quasiparticle model
from the $t$-$J$ model has been developed \cite{bel97}. It used a 
generalization of the canonical transformation (CT) approach of the 
Lang-Firsov type. An effective Hamiltonian for the spin polarons 
includes in itself both types of the hole-hole interactions mentioned
above in a natural way. Some details of this approach
are presented in Section IV. 
In Ref. \cite {bel97} results for the single-hole properties have
been compared with ones of the SCBA calculations and an impressive
agreement has been found.  This is supported further by the idea that the
``canonically transformed" quasiparticles are close to exact $t-J$
model ones.   
Even though CT approach is less controlled than the SCBA one, it solves 
naturally all the problems mentioned above and allows one to derive the 
quasiparticle Hamiltonian for interacting spin-polarons from the
original $t$-$J$ model. Thus,
in this paper we compare the predictions obtained from this Hamiltonian
to our numerics, and in this way we critique the description of
the interactions between quasiparticles that follows from the CT approach.

\subsection{Numerical studies}

ED studies of the $t-J$ model doped away from half-filling 
on small clusters with periodic boundary conditions are
an important source of unbiased information on the
low-energy physics of this system. One- and two-hole ground states
have been investigated in great detail on the 16- ($4\times 4$) 
\cite{Prel2,PD1,Riera,HP,Dag,DRY,Riera2,Ding,Barnes}, 18-
($\sqrt{18}\times\sqrt{18}$), 20- ($\sqrt{20}\times\sqrt{20}$) 
\cite{Elser,Sz,Itoh,Fehske,SH,EdOh1,Ed_other,EdOh2,EdOh3,EdOh4}, and
26-site ($\sqrt{26}\times\sqrt{26}$)
\cite{Poil2,Poil3,Poil4,Prel1,Poil1,p94,prd94} clusters. 
Although some of these results converge, at least partially, these
clusters suffer from various finite size problems.
The 20- and 26-site clusters do not have the full rotational
symmetry of the square lattice. Therefore, they do
not possess important reciprocal lattice points along the high symmetry directions
in the first Brillouin zone.
For example, the important reciprocal lattice point $(\pi/2,\pi/2)$ 
does not exist in the first Brillouin zone of the 
18-, 20-, and 26-site clusters.  This causes the  ground state momenta of 
the one-hole state to be different from the predicted
$\pm(\pi/2,\pm\pi/2)$ points. 
Although the 16-site cluster has the $(\pi/2,\pi/2)$ point, 
it has an additional symmetry
which causes an accidental degeneracy of the levels at $(\pi/2,\pi/2)$ and
$(\pi,0)$ for one hole, and between $(0,0)$ and $(\pi,0)$ for the
two-hole problem \cite{Barnes}. Attempts to remedy the missing $(\pi/2,\pi/2)$ 
point have been made by using
the non-square 16-site ($\sqrt{8}\times\sqrt{32}$) \cite{G1} and
24-site ($\sqrt{18}\times\sqrt{32}$) \cite{VGLC} clusters. 

Previous results on the single-hole problem
show that the quasiparticle peak at the bottom of the
spectral function can be expected to survive in the thermodynamic limit 
\cite{lg95,PD1,Dag,Poil3}. The corresponding quasiparticle band is
narrow (of the order of $2J$ in the ``physical'' region $J/t<1$)
and the band minima are shifted to the ABZ boundary. However, 
due to the previously mentioned deficiencies of the
16-, 18-, 20-, and 26-site clusters, none of them
can actually show that the quasiparticle minima are at 
the $\pm(\pi/2,\pm\pi/2)$ points; these wave vectors are
those predicted by numerous theoretical studies \cite{KLR,SVR,Trug,SS2}. 

\begin{figure}
\unitlength1cm
\epsfxsize=7cm
\begin{picture}(4,9)
\put(1.5,1.5){\rotate[r]{\epsffile{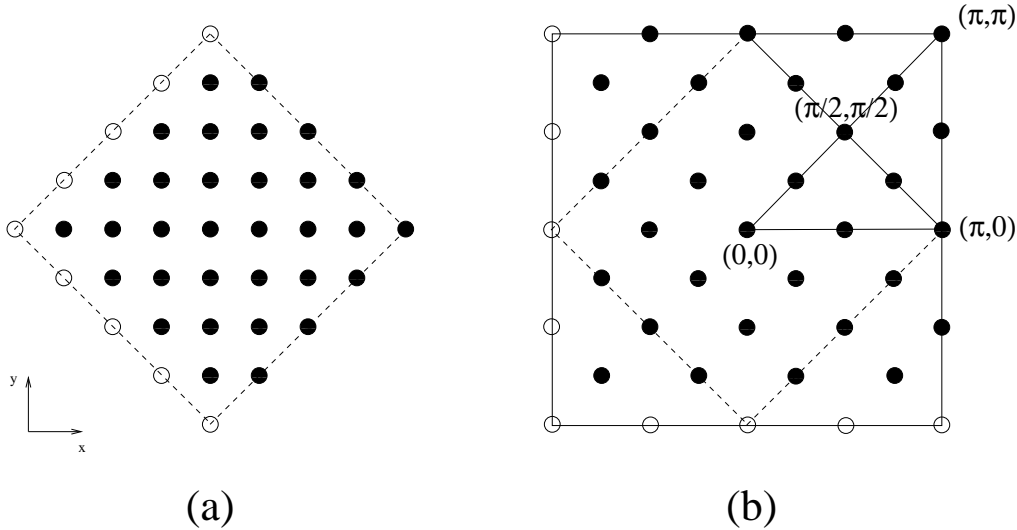}}}
\end{picture}
\caption{ The 32-site cluster in (a) direct and (b) reciprocal space. Empty 
circles are included to mimic the periodic boundary conditions used in our
studies. The solid lines in (b) show the important directions in 
${\bf k}$-space displaying the high symmetry of the cluster. The dashed 
line in (b) borders the first magnetic Brillouin zone.}
\label{cluster}
\end{figure}

The smallest cluster which has the full rotational symmetry of the square
lattice, contains the $(\pi/2,\pi/2)$ point, and is free from the spurious 
degeneracies mentioned above, is the 32-site cluster 
($\sqrt{32}\times\sqrt{32}$)  --- see Fig.\ref{cluster}. 
Also, it is the largest such system which
can be solved using modern computers. These calculations involve finding
the lowest energy states of matrices with dimensions of up to
300 million.

Recently, some results for the single-hole 
problem have been published for this cluster by two of us \cite{lg95}.
These numerics showed that the effective mass around the minima is
anisotropic, and that the quasiparticle residue is rather small for 
realistic $J/t$, both in excellent agreement with analytical
predictions. Further, the full dispersion relation predicted by
analytical work based on the SCBA \cite{Man1,Horsch,FM} 
is found to be in excellent agreement with ED numerics on this
cluster \cite{lg95}.

ED results for two holes on finite clusters consistently show that they are 
coupled in a bound state with $d_{x^2-y^2}$ symmetry in a wide range of $J/t$
\cite{Prel2,Riera,HP,DRY,Riera2,Barnes,Itoh,Fehske,EdOh3,Poil1,p94,prd94,Prel},
in agreement with earlier ED data for the 16-site Hubbard model
\cite{Kax,Riera1}, the Green's Function Monte Carlo studies on $8\times
8$ cluster \cite{bm93}, and some theories
\cite{Ed,bel97,SK,SWZ,FKS,NazD,Hirsch}. 
Low lying states with other symmetries, as
well as ${\bf k}\neq (0,0)$ states, have also been studied \cite{DRY,EdOh3}.
Attempts have been made 
to extrapolate the binding energy to the thermodynamic limit
and thus to estimate the critical value of $J/t$ for the formation of a bound
state \cite{Poil1,prd94}. Also, some knowledge concerning the
internal structure of the bound state is known \cite {p94,bm93}.
Lastly, the electron momentum distribution function has
been investigated in order to search for Fermi-like 
discontinuities. Drastic differences between the single-hole and
two-hole cases have been noted \cite{Ding,SH,EdOh1}. Finite size and
$J/t$ scalings of this quantity have also been studied
\cite{EdOh1,Ed_other,EdOh4}. 

Nevertheless, the above-mentioned results are of limited usefulness simply 
because of the systematic error introduced by the lower symmetry of the 
clusters with 16, 18, 20, and 26 sites. Clearly, the 32-site cluster
would augment such studies. Further, with the collection of all such clusters,
and some analytical guidance regarding the correct scaling laws,
information on the thermodynamic limit, {\it viz.} a density of
zero (two holes in an infinite 2D square lattice), would be accessible.

Section III will summarize the results of the ED studies of
the single- and two-hole problems that we have obtained on the 32-site cluster. 
Some of the single-hole results have been published previously \cite{lg95},
and we only mention those results that are crucial to our scaling analyses. 
A brief summary of a portion of the two-hole comparison to the CT Hamiltonian
was presented in Ref. \cite{clg98}. 

\section{Numerical work}

Most of the results presented in this section are obtained by ED on the 32-site
cluster with periodic boundary conditions for the realistic value $J/t=0.3$. 
Results at different $J/t$, as well as on smaller clusters published previously 
\cite{Ding,Poil1}, are also used in the discussion of the finite size scaling 
(FSS), bound state energies, and correlation functions.

To look for evidence of hole binding in the low energy states, we calculate 
the two--hole binding energy $E_b \equiv E_2-2E_1+E_0$, where $E_1$ and 
$E_0$ are the ground state energies with one and no hole respectively, and 
$E_2$ is the energy of the two-hole state.
Further, two expectation values that we are interested in are defined as 
follows: (i) The electron momentum distribution function (EMDF) is given by
$\langle n_{{\bf k}\sigma} \rangle \equiv
\langle \tilde{c}^\dagger_{{\bf k}\sigma}
\tilde{c}_{{\bf k}\sigma}\rangle$, where $\tilde{c}^\dagger_{{\bf
k}\sigma}$, $\tilde{c}_{{\bf k}\sigma}$
are the Fourier transform of the constrained operators.
(ii) The spatial distribution of holes in the bound state is characterized
by the pair correlation function defined as 
\begin{equation}
C(r)=\frac{1}{N_hN_E(r)}\sum_{i,j}
\langle(1-n_i)(1-n_j)\delta_{|i-j|,r}\rangle,
\label{eq:cofr}
\end{equation}
where $N_h$ is the number of holes, and $N_E(r)$
is the number of equivalent sites at a distance $r$ from any given site.

Before presenting the FSS of the EMDF, we discuss what kind of finite size 
behaviour one can expect. The EMDF is expected to show how hole
doping changes the uniform value of $\langle n_{{\bf k}\sigma} \rangle
=\frac{1}{2}$ obtained in the half-filled case. In a system of
free particles, a hole with a certain momentum will manifest itself
as the complete suppression of $\langle n_{{\bf k}\sigma} \rangle$ 
to zero at this momentum. In systems with interaction whose physics can
be described in terms of the quasiparticles, this 
suppression will be proportional to the quasiparticle residue $Z_{\bf k}$,
and is almost independent of the cluster size; the rest of the hole weight
will be distributed among the other available ${\bf k}$-points.
Consequently, the more ${\bf k}$-points a system
possesses the less hole weight each ${\bf k}$-point will carry. 
Therefore, in the single-hole problem
we anticipate that $\langle n_{{\bf k}\sigma} \rangle$
for the ground state momentum ${\bf P}$ to be suppressed by a constant
proportional to $Z_{\bf P}$.  
Further, we expect that the deviation from the half-filled value,
\begin{equation}
  \langle\delta n_{{\bf
k}\sigma}\rangle= \langle n_{{\bf k}\sigma}\rangle -\frac{1}{2},
\label{eq:delta_n}
\end{equation}
will scale as $1/N$ at all other points until (roughly) the physics of 
the system does not change with
the cluster size, {\it i.e.}, when size of the
quasiparticle is smaller than the cluster size. 

The same argument should apply to the bound states of the two-hole problem, 
where we predict $\langle\delta n_{{\bf k}\sigma}\rangle$ to scale as $1/N$ 
at all ${\bf k}$-points. 

\subsection{Single-hole case}

We wish to provide a FSS analysis of certain quantities for the two-hole 
ground state. To this end, we present new results for the one-hole 
problem that will facilitate such work.

\begin{figure}
\unitlength1cm
\epsfxsize=7cm
\begin{picture}(4,9)
\put(0.0,0.5){\epsffile{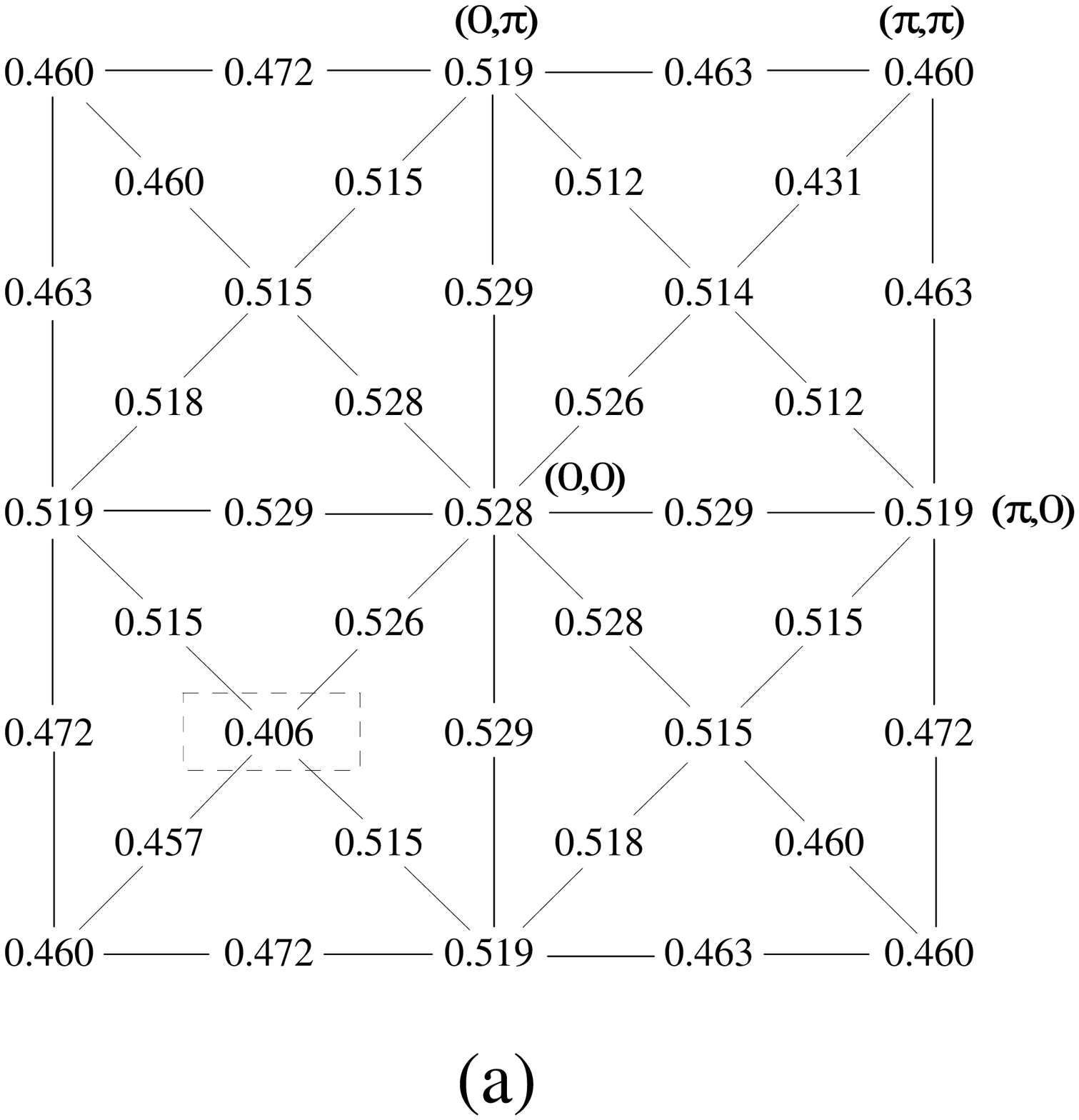}}
\end{picture}
\unitlength1cm
\epsfxsize=7cm
\begin{picture}(4,9)
\put(6.0,0.5){\epsffile{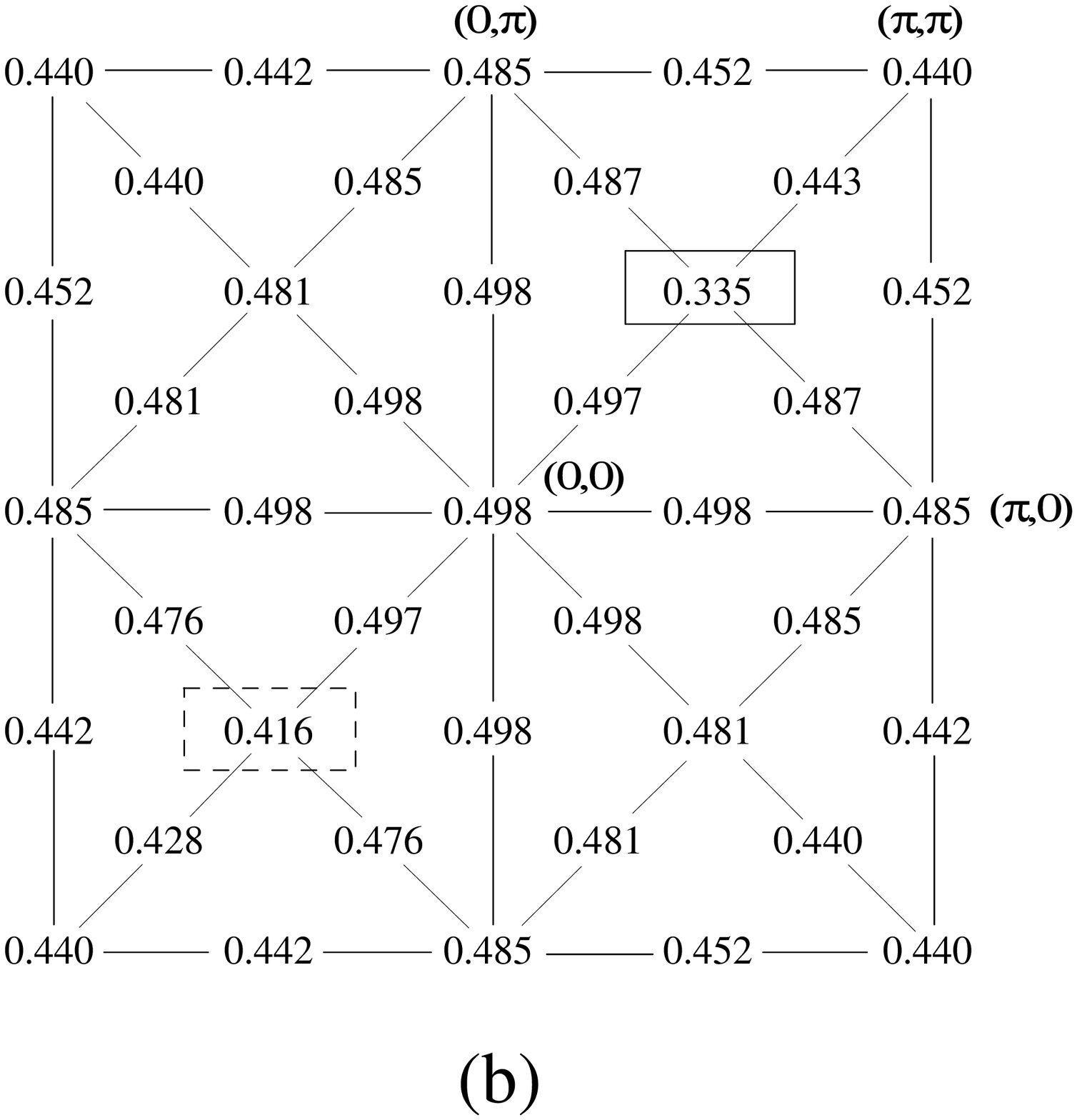}}
\end{picture}
\caption{The EMDF for the single-hole ground state having momentum ${\bf
P}=(\pi/2,\pi/2)$ and $S^z_{tot}=+1/2$ at $J/t=0.3$. The numbers are
the electron filling factors (a) $\langle n_{{\bf k}\uparrow}\rangle$, 
(b) $\langle n_{{\bf k}\downarrow}\rangle$ at the corresponding ${\bf
k}$-points. ``Antidips'' at $(-\pi/2,-\pi/2)$ described in the text 
are denoted by the dashed boxes, and the ``dip'' in (b) at the ground-state 
momentum is highlighted by the solid box.}
\label{nqupdown}
\end{figure}

Figure~\ref{nqupdown} shows the
EMDF of the single-hole ground
state on the 32-site cluster at $J/t=0.3$, which has
total spin $S^z_{tot}=+\frac{1}{2}$ and momentum 
${\bf P}=(\pi/2,\pi/2)$. Due to the non-zero momentum of this state, the
only symmetry its EMDF has is a reflection about
the ``main diagonal'' ($(-\pi,-\pi)\leftrightarrow (\pi,\pi)$ line).

Several features of the EMDF are worth noticing. 
First, $\langle n_{{\bf k}\downarrow}\rangle$ 
has a ``dip'' at the GS momentum ${\bf P}$.
Earlier studies of the $J/t$
dependence of the intensity of this ``dip'' have left no doubt about
its direct relation to the quasiparticle weight 
$Z_{\bf P}$ \cite{EdOh1}. Second,
the EMDF deviates significantly
from its half-filled value for both
spin directions across the 
entire Brillouin zone. This background has
a maximum at $(0,0)$ and a minimum at $(\pi,\pi)$, which is a biproduct
of minimizing the kinetic energy of the system \cite{EW}. 
Although this ``dome'' shape resembles the ``large Fermi surface'' in
a system of free electrons, it has different physics behind it. The
discussion of this behaviour will be given in \S IV.
One observes that this dome
structure in $\langle n_{{\bf k}\uparrow}\rangle$ is shifted
upwards from its half-filled value, and that in
$\langle n_{{\bf k}\downarrow}\rangle$ it is shifted downwards ($\langle
n_{(0,0)\uparrow}\rangle - \langle n_{(0,0)\downarrow}\rangle\simeq
0.03$).  The difference between the maximum and minimum,
\begin{equation}
\Delta n_{\sigma}= \langle n_{(0,0)\sigma}\rangle -\langle
n_{(\pi,\pi)\sigma}\rangle,
\end{equation}  
is slightly larger for 
$\sigma=\uparrow$ than for $\downarrow$ ($\Delta
n_{\uparrow(\downarrow)}\simeq 0.07 (0.06)$ at $J/t=0.3$).
Also $\Delta n_{\downarrow}$ has a stronger $J/t$ dependence. 
Finally, the EMDFs of both $\sigma=\uparrow,\downarrow$
have ``antidips''  at 
$(-\pi/2,-\pi/2)$ . They
were observed earlier but no successful explanation has been
presented. The fact that the ``antidips'' are always at ${\bf P}-{\bf Q}_{AF}$
supports the idea that their physics is somehow related to the
long-range AF fluctuations in the system \cite{EdOh1}. 
Subtraction of the ``normal'' background from $\langle
n_{(-\pi/2,-\pi/2)\sigma}\rangle$ shows
that the depth of the ``antidip'',
\begin{equation}
\Delta n_{anti,\sigma}= \langle
n_{(-\pi/2,-\pi/2)\sigma}\rangle-\langle n_{(\pi/2,-\pi/2)\sigma}\rangle,
\end{equation}
is larger for $\uparrow$ ($\Delta n_{anti,\uparrow(\downarrow)}\simeq 0.11
(0.08)$ at $J/t=0.3$), and $\Delta
n_{anti,\downarrow}$ has stronger $J/t$ dependence.

\begin{figure}
\unitlength1cm
\epsfxsize=6cm
\begin{picture}(4,9)
\put(0.5,0){\epsffile{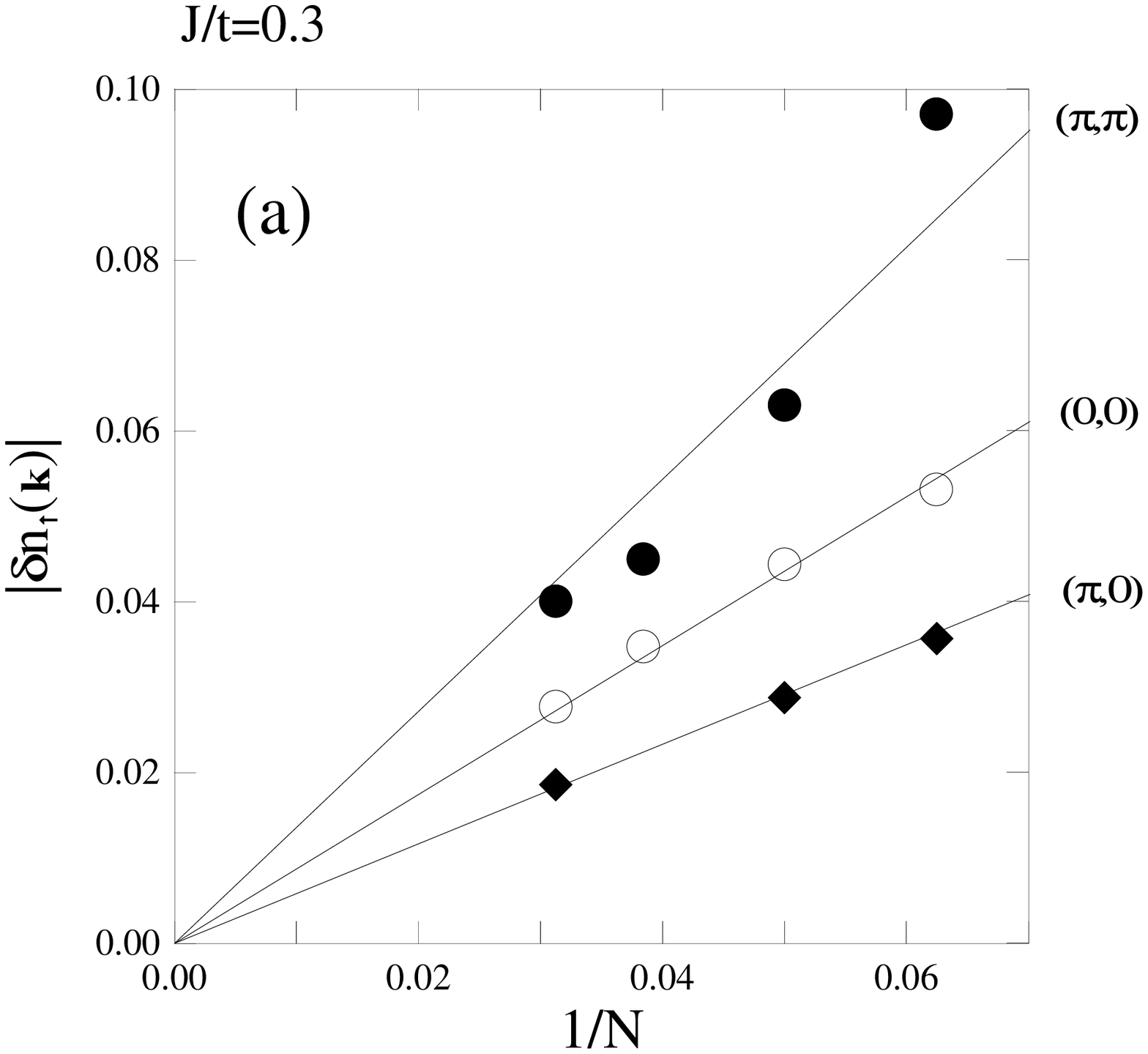}}
\end{picture}
\unitlength1cm
\epsfxsize=6cm
\begin{picture}(4,9)
\put(5.5,0){\epsffile{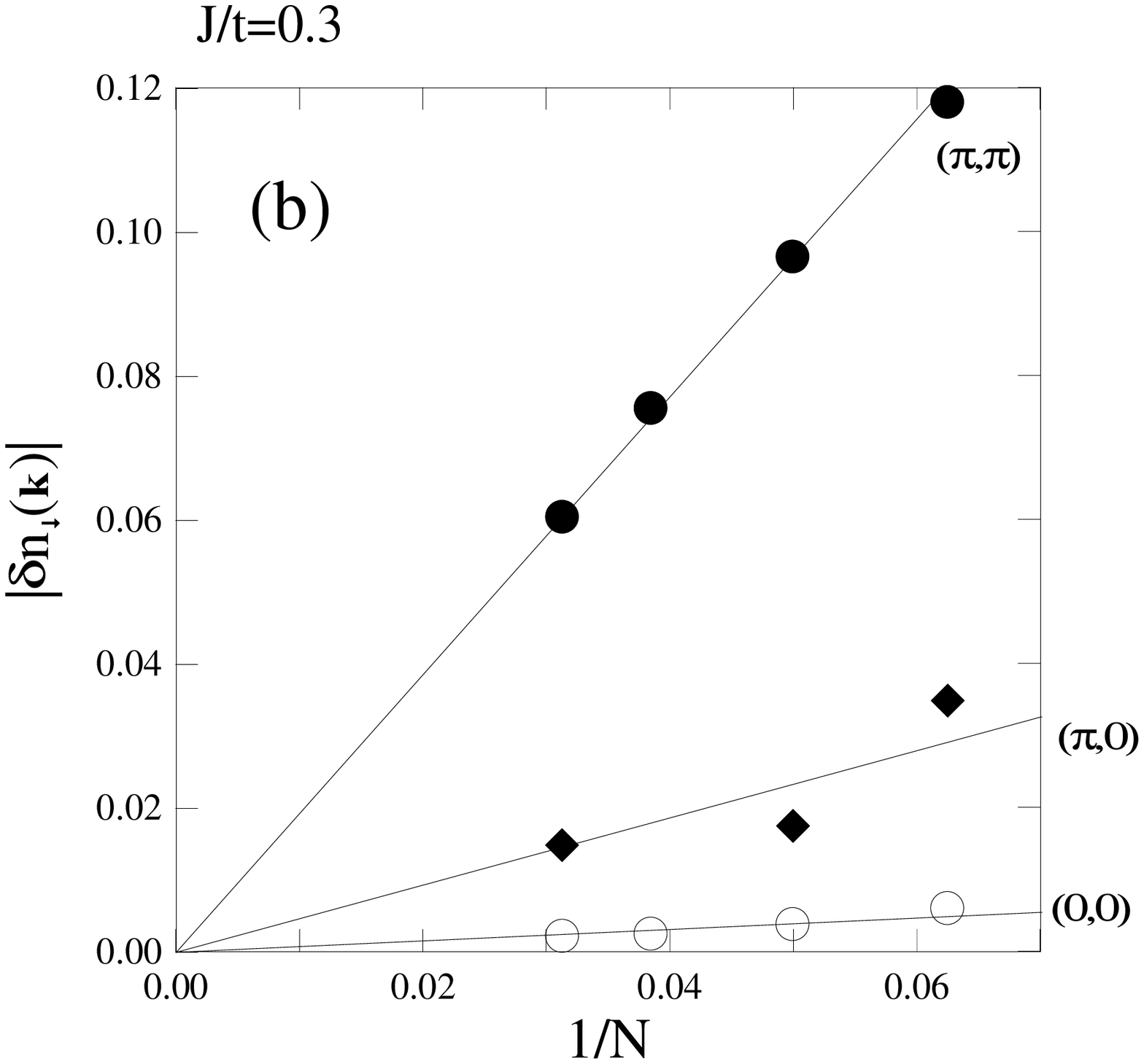}}
\end{picture}
\caption{$1/N$ scaling of the  
(a) $|\langle\delta n_{{\bf k}\uparrow}\rangle|$, 
(b) $|\langle\delta n_{{\bf k}\downarrow}\rangle|$ 
(see Eq.~(\ref{eq:delta_n})) 
for ${\bf k}=(\pi,\pi)$ (filled circles), $(0,0)$ (empty circles), and
$(\pi,0)$ (filled diamonds) for the single-hole ground state 
at $J/t=0.3$. The data from 16-, 20-, 26-, and
32-site clusters ($(\pi,\pi)$, $(0,0)$ points) and from 16-, 20-, and
32-site clusters ($(\pi,0)$ point), where these points are available, 
are used.}
\label{n1_03a}
\end{figure}

\begin{figure}
\unitlength1cm
\epsfxsize=6cm
\begin{picture}(4,9)
\put(0.5,0){\epsffile{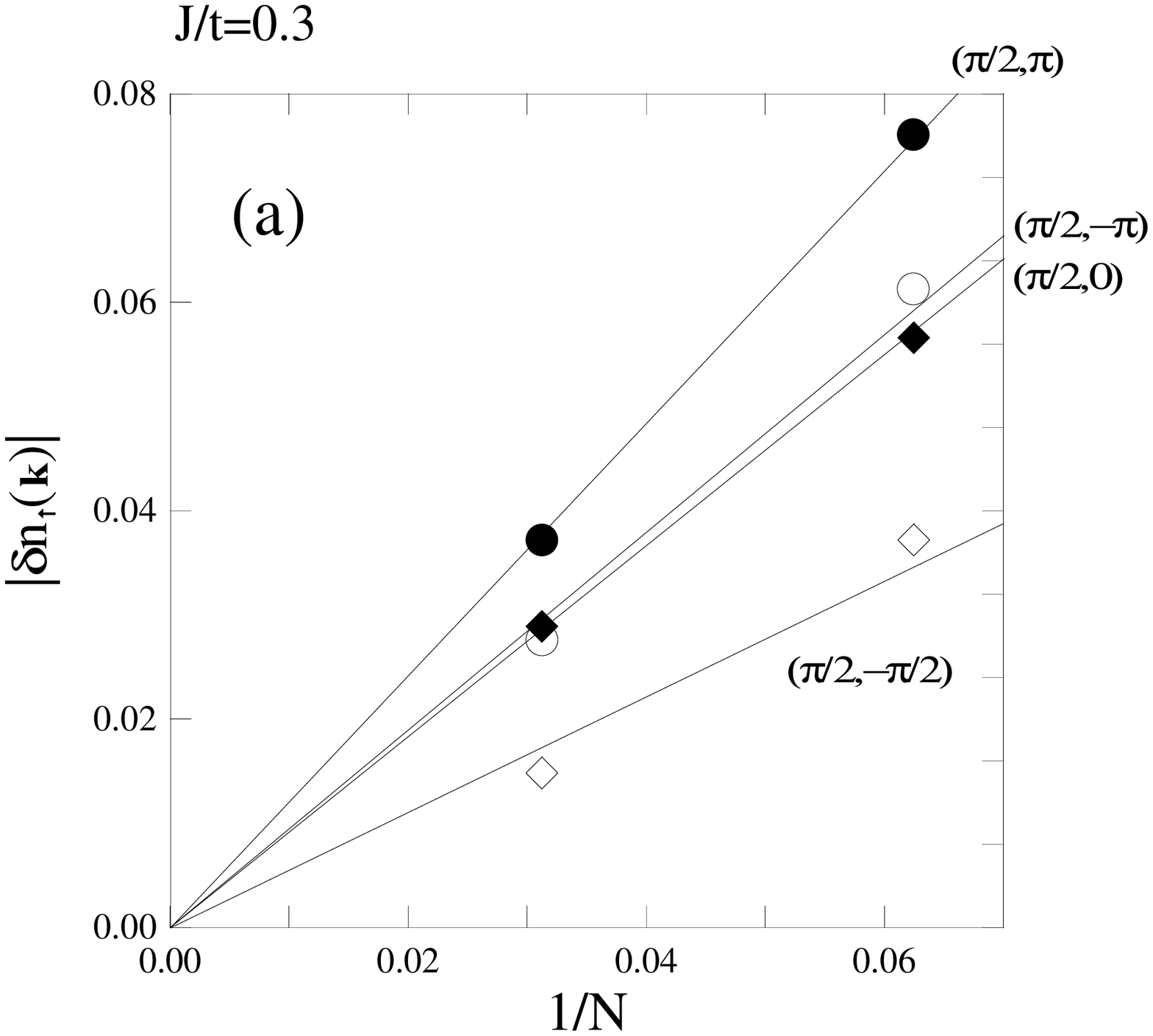}}
\end{picture}
\unitlength1cm
\epsfxsize=6cm
\begin{picture}(4,9)
\put(5.5,0){\epsffile{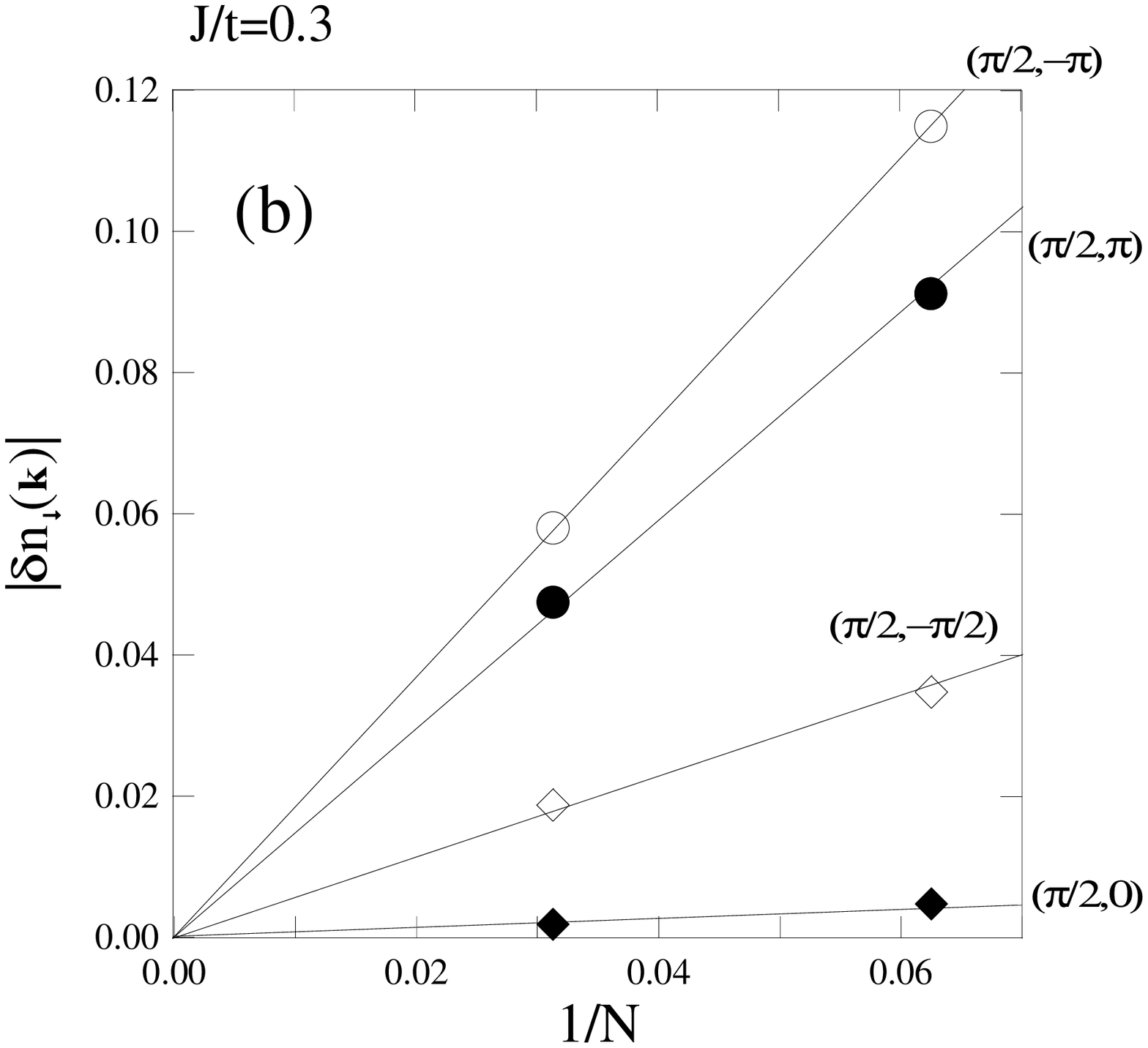}}
\end{picture}
\caption{The same as in Fig.~\ref{n1_03a} for ${\bf k}=(\pi/2,\pi)$ 
(filled circles), $(\pi/2,-\pi)$ (empty circles), $(\pi/2,0)$ (filled
diamonds), and $(\pi/2,-\pi/2)$ (empty diamonds).
These points are available from 16- and 32-site clusters only.}
\label{n1_03b}
\end{figure}

\begin{figure}
\unitlength1cm
\epsfxsize=6cm
\begin{picture}(4,9)
\put(5,0){\epsffile{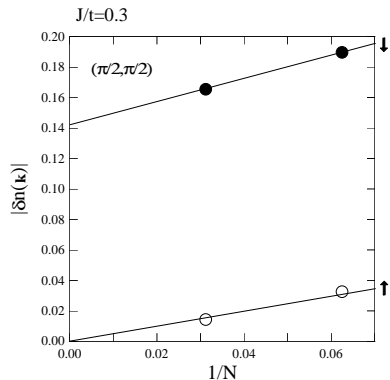}}
\end{picture}
\caption{$1/N$ scaling for the 
$|\langle\delta n_{{\bf k}\uparrow}\rangle|$ (open
circles) and $(C+\alpha/N)$ scaling for the ``dip'' in  $|\langle\delta
n_{{\bf k}\downarrow}\rangle|$ 
(filled circles) at the ground-state momentum
${\bf k}={\bf P}$, $J/t=0.3$.
These points are available for the 16- and 32-site clusters only.}
\label{n1spec}
\end{figure}

In Figs. \ref{n1_03a} and \ref{n1_03b} we plot 
$\langle\delta n_{\sigma{\bf k}}\rangle$
vs. the
``inverse volume" of the system, $1/N$, at $J/t=0.3$ for
all ${\bf k}$-points (except ${\bf P}$ and ${\bf P-Q}_{AF}$) available on
more than one cluster.
One can see the almost perfect $1/N$ scaling 
at all these points, in agreement with our
expectation. 
Figure~\ref{n1spec} shows the same plot 
for the EMDF at the ground state momentum. 
Extrapolation to the thermodynamic
limit  shows that the dip, which we expect to be $Z_{\bf P}/2$
(the factor one-half is from the
proper normalization of the wave function),
 is about $0.14$, or $Z_{\bf P} \simeq 0.28$.
This agrees well with SCBA result, $Z_{\bf P}^{SCBA}=0.271$ 
\cite{Man1}. There is no simple scaling for the ``antidips'' of
$|\langle\delta n_{{\bf P-Q}\sigma}\rangle|$
 because of the long-range physics involved. 
According to the discussion in Sec. IV they are
combinations of terms with different scaling behaviours.

\subsection{Two-hole case}

\begin{figure}
\unitlength1cm
\epsfxsize=6cm
\begin{picture}(4,7.5)
\put(0,1.5){\rotate[r]{\epsffile{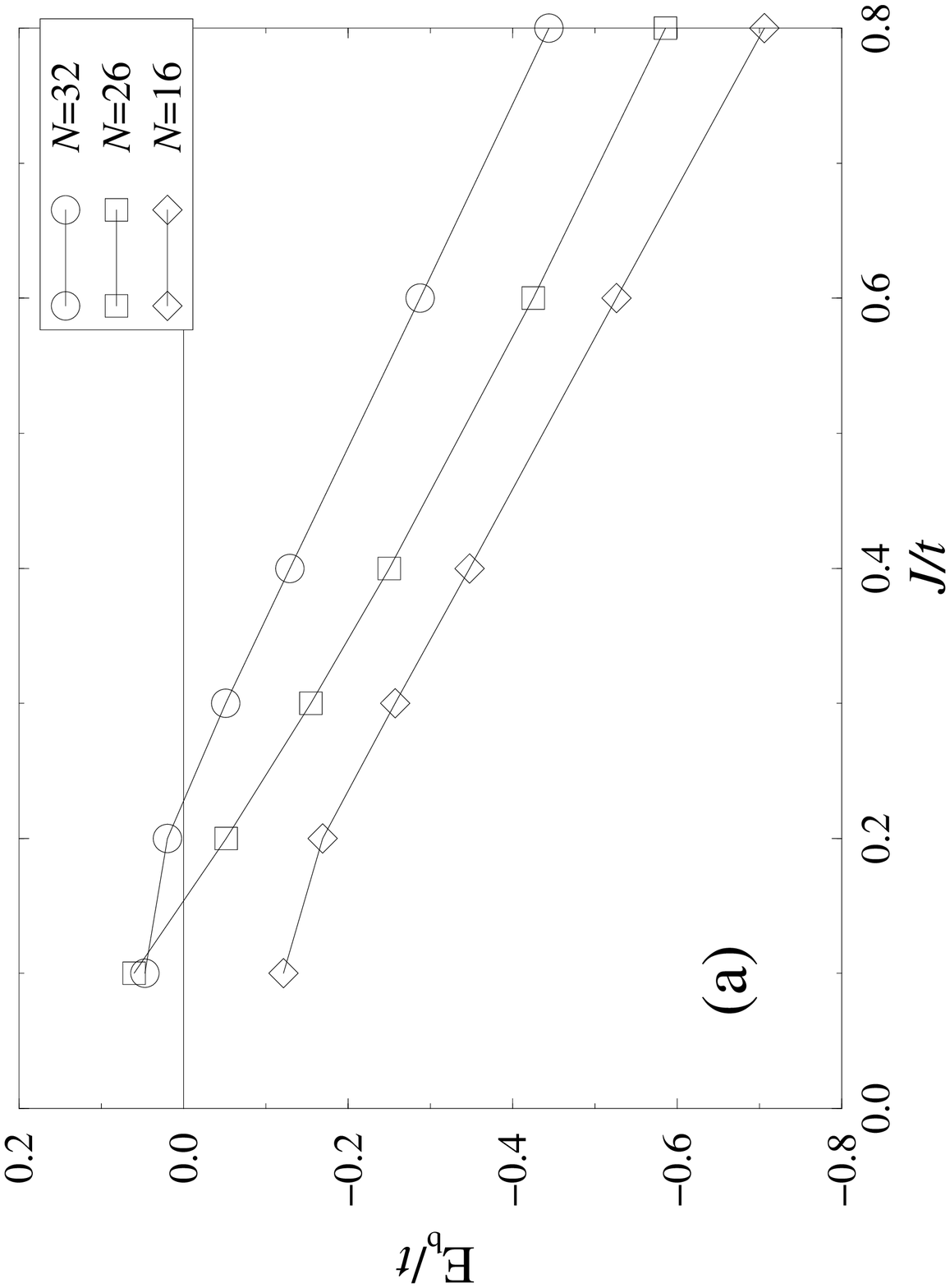}}}
\end{picture}
\unitlength1cm
\epsfxsize=6cm
\begin{picture}(4,7.5)
\put(6.0,0){\epsffile{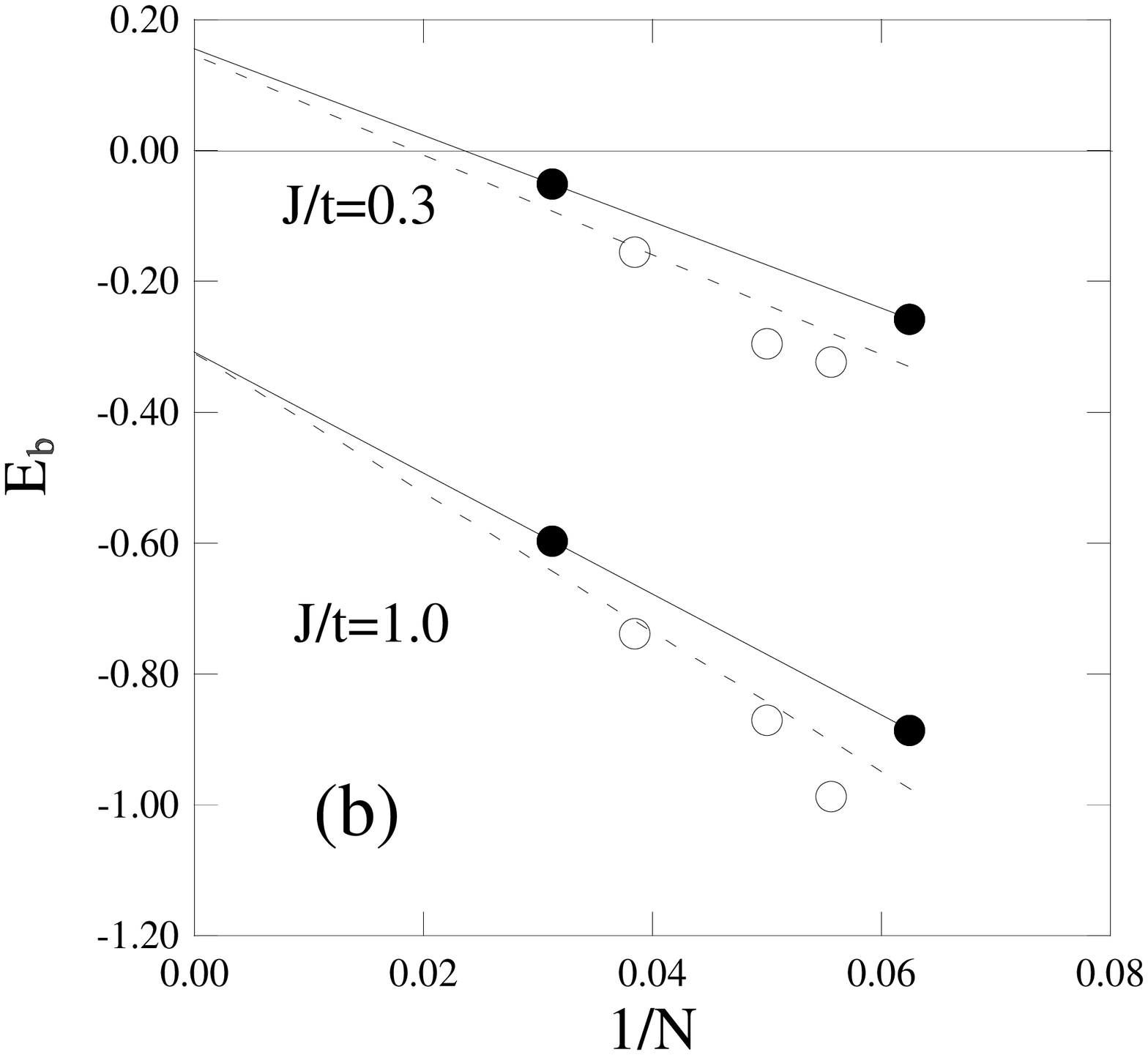}}
\end{picture}
\caption{(a) The $J/t$ dependence of the binding energy $E_b$, in units
of $t$, from ED studies on the 16- (diamonds), 
26- (squares), and 32-site (circles) 
clusters.  (b) The binding energy {\it vs.} 
$1/N$ for two representative $J/t$ values, 
$J/t=0.3$ (upper) and $J/t=1.0$ (lower). The solid and dashed lines are $1/N$
scaling using 16- and 32-site data only (filled circles) and data from
all available clusters (open and filled circles), respectively.}  
\label{En_2}
\end{figure}

The only zero total momentum bound state that we have found in
the zero magnetization channel is a singlet and it has $d_{x^2-y^2}$
symmetry.  
Figure~\ref{En_2}(a)  shows the $J/t$
dependence of the binding energy $E_b$ on the 16-, 26-, and 32-site
clusters.
One can see that the absolute value of the
binding energy gets smaller as the size of the cluster grows,
and that in some region of $J/t$ the binding energy becomes positive.
Such behaviour seems to be natural in the presence of short-range
attraction between the holes. In this case holes on larger clusters 
lower their kinetic energy due to delocalization and make the bound
state shallower, whereas on smaller ones they are not allowed to move 
farther apart. Further, holes on smaller clusters are forced to be in
the region of the mutual attraction.
 Since these short-range interactions are believed to be of
magnetic origin,  the interaction energy has to scale as $J$. 
Consequently, the increasing importance of the kinetic energy at small
$J/t$ tends to destroy the bound state. 
This line of thinking leads to a discussion of whether or not
the critical threshold of $J/t$ for bound state
formation is above or below the ``realistic" value of $J/t$ for
the cuprates. Attempts have been made to estimate the thermodynamic limit of 
$(J/t)|_c$ through FSS of the binding energy
\cite{Poil1,prd94}. If we follow the same approach, we obtain the scaling 
shown in Fig.~\ref{En_2}(b), and
this data shows the FSS at two
representative $J/t$ values. The thermodynamic limit of $E_b$ is negative
at the larger $J/t$ (smaller size of the bound state, larger role of
the short-range interaction) and positive at the smaller
$J/t$ (no bound state).

In our opinion, this approach is problematic because for at
least two reasons. First, there is another important hole-hole
interaction, {\it viz.} magnon exchange, which also leads to
pairing. In fact, it is this interaction which selects the $d$-wave
symmetry state. It has been shown analytically
\cite{bel97,SS2,SK} that this interaction alone leads to the
formation of a shallow long-range bound state which does not
have a critical value of $J/t$ because the interaction strength grows
with $t$.  
Therefore, one can expects that further increase in the cluster size will 
not only lower the kinetic energy of the holes, but will also provide 
more sites for the holes to take advantage of the long-range attraction. 
The second reason is the
absence of the evident scaling law for the binding energy.  
Regarding the complexity of the interactions involved and the tendency
of the magnetic subsystem towards AF long-range order,
we expect different contributions to the FSS of $E_b$
which are of different order in $1/N$ and of
comparable magnitudes.
This is especially true at smaller $J/t$ when the size
of the bound state is comparable to or larger than the cluster size.

\begin{figure}
\unitlength1cm
\epsfxsize=6cm
\begin{picture}(8,7)
\put(-0.3,1){\rotate[r]{\epsffile{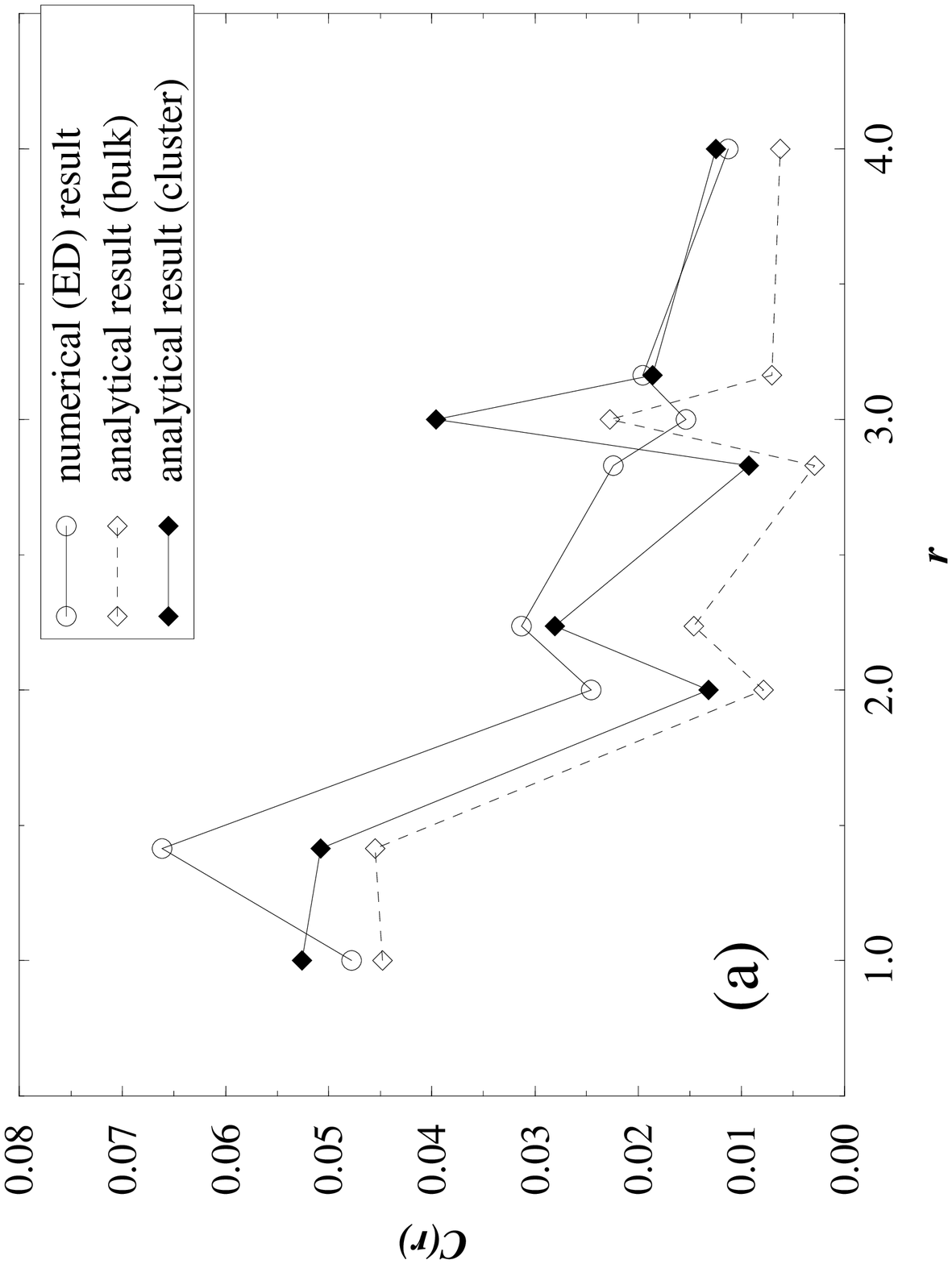}}}
\end{picture}
\unitlength1cm
\epsfxsize=6cm
\begin{picture}(8,7)
\put(1.3,1){\rotate[r]{\epsffile{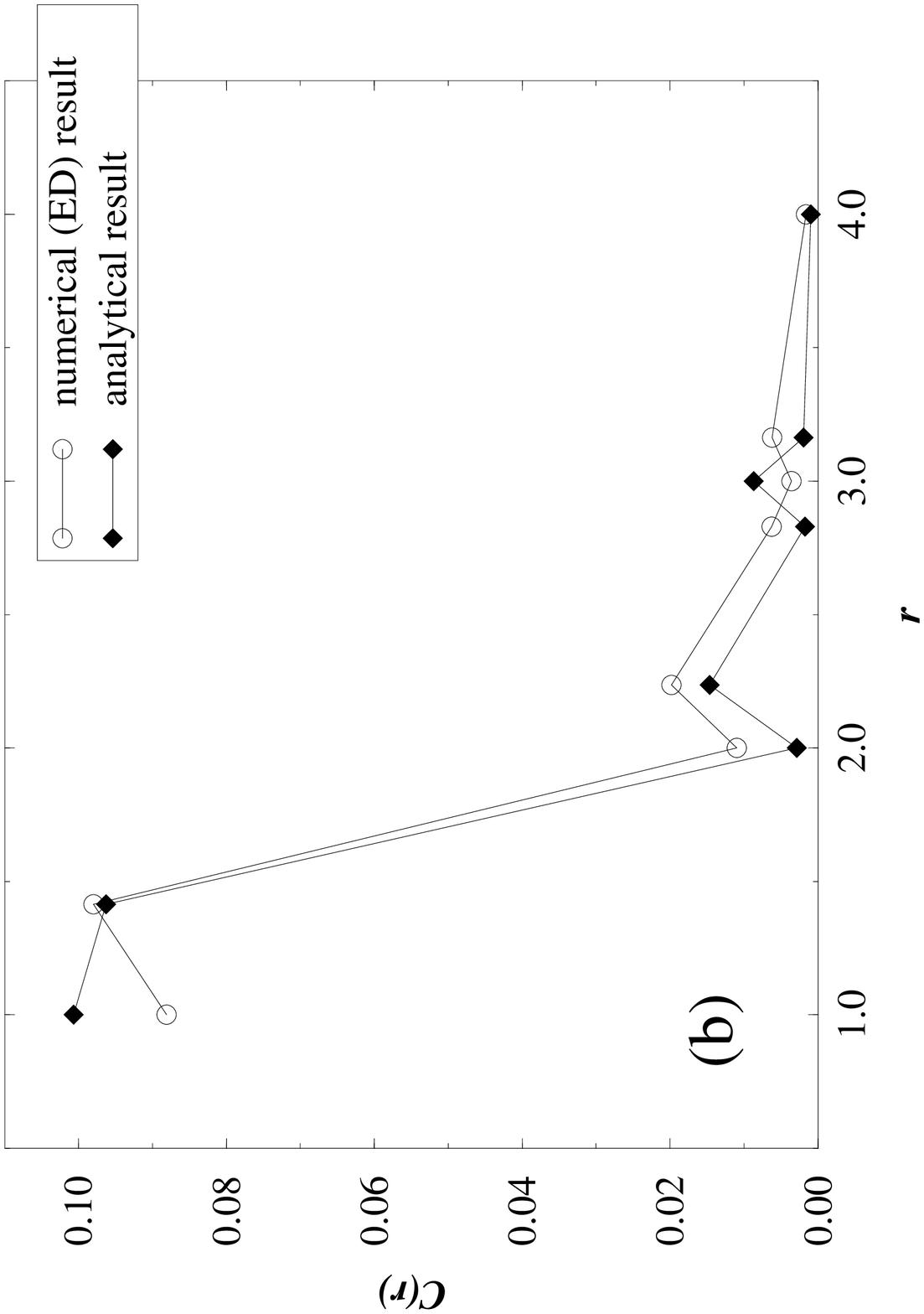}}}
\end{picture}
\caption{The spatial correlation function, $C(r)$, for two holes doped
into a square lattice described by the $t$-$J$ model, for (a) $J/t =
0.3$, and (b) $J/t=1.0$.
Our ED results (open circles), the analytical results for 
an infinite square lattice (open diamonds), 
and the analytical results mapped onto a 32-site square lattice (filled
diamonds), are all shown. The lines are a guides to the eye.
In (b), analytical results for the cluster are
very close to ones for the bulk, and hence are not shown.} 
\label{C_r}
\end{figure}

Another important quantity which shows further evidence of the hole-hole
attraction in an AF background is the hole-hole correlation function
$C(r)$, Eq. (\ref{eq:cofr}). It has been studied in detail on
smaller systems \cite{Prel2,Riera2,Poil1,prd94,Prel}. Figures~\ref{C_r}(a)
and \ref{C_r}(b) show the 32-site ED results for $C(r)$ at $J/t=0.3$ and
$J/t=0.8$, respectively. In a wide region of $J/t$ the strongest
correlation is at the $\sqrt{2}$ distance, while the nearest-neighbour
correlation is also strong. 
A density-matrix renormalization group study \cite {w96} has
also found similar physics. At larger $J/t$ (Fig. \ref{C_r}(b)) 
the size of the bound state is small:
the nearest neighbour and $\sqrt{2}$ distances accumulate about 80\% of 
the holes. However, at $J/t=0.3$ the {\it probabilities} of finding the
holes at $\sqrt{5}$ and $\sqrt{2}$ distances are
almost the same, and only 46\% of the holes are located at
the nearest neighbour and $\sqrt{2}$ distances.
The correlation decays slowly with distance at small
$J/t$. Hence in the $J/t=0.3$ bound state one can expect
$C(r)$ to have a longer ``tail'' in the thermodynamic limit.
\begin{figure}
\unitlength1cm
\epsfxsize=7cm
\begin{picture}(8,9)
\put(0,0.5){\epsffile{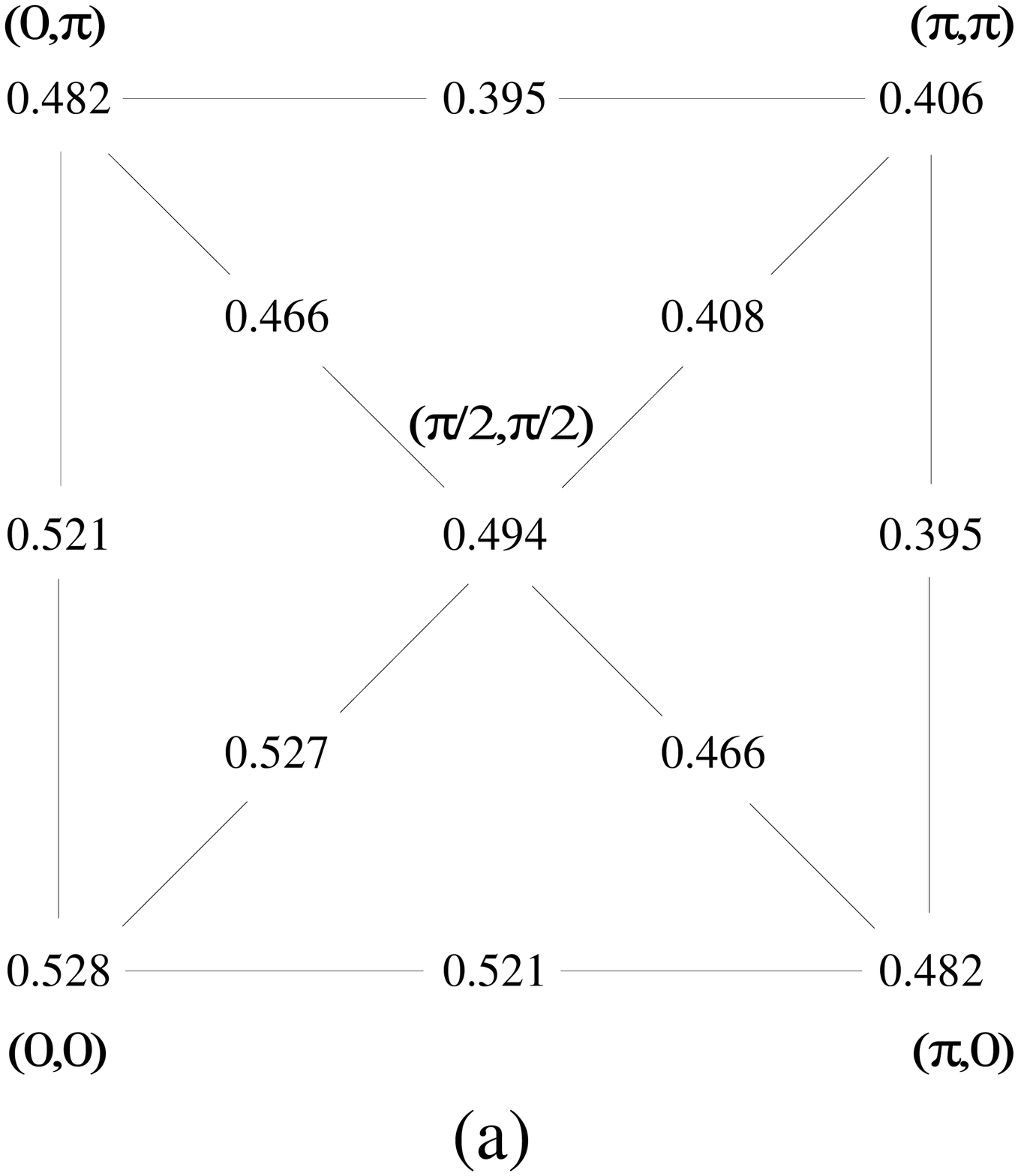}}
\end{picture}
\unitlength1cm
\epsfxsize=7cm
\begin{picture}(8,9)
\put(2,0.5){\epsffile{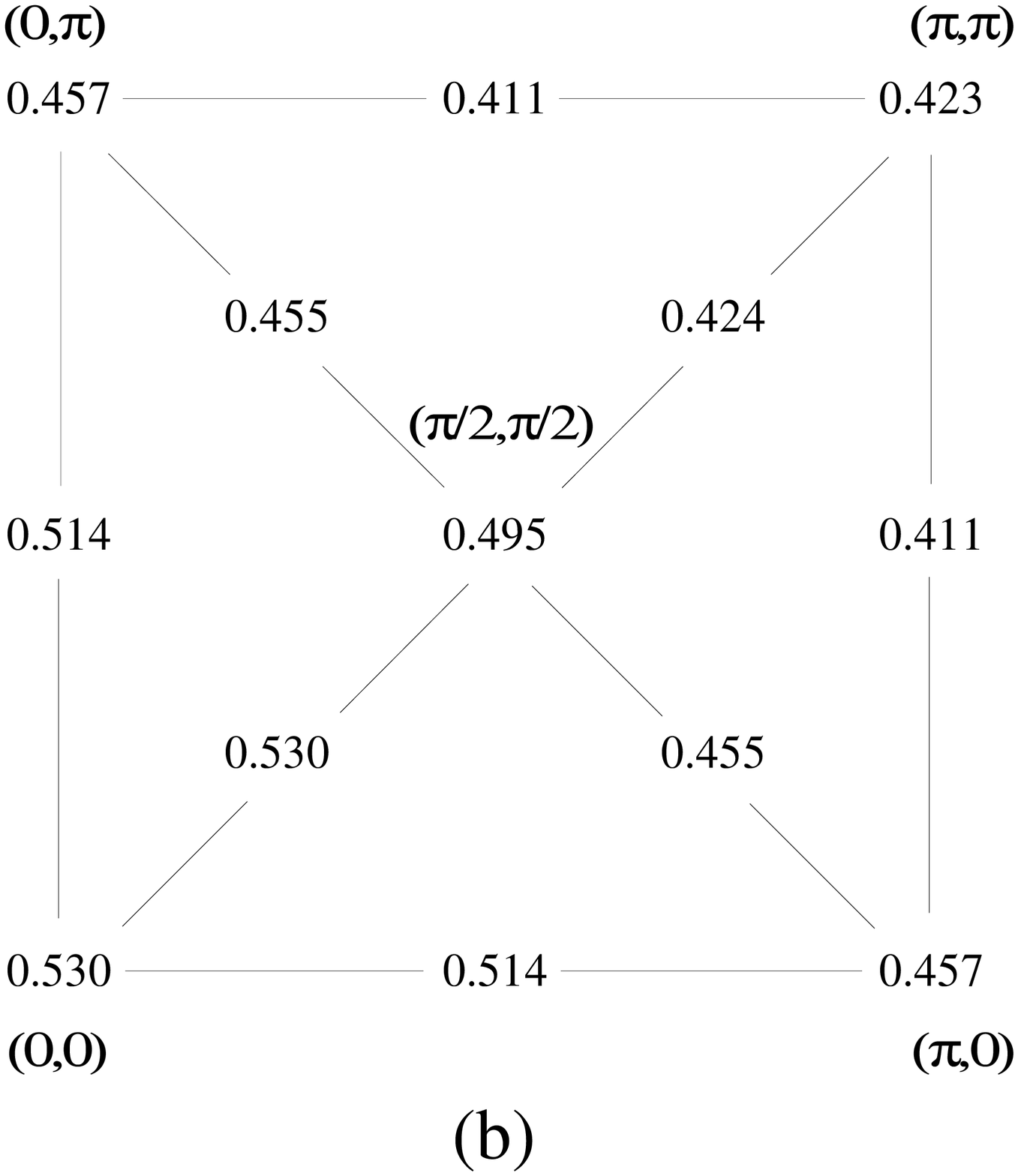}}
\end{picture}
\caption{The EMDF for the two-hole ground state, ${\bf P}=0$,
$S^z_{tot}=0$ at (a) $J/t=0.3$, (b) $J/t=1.0$ within the first
quadrant of the BZ.}
\label{nq2}
\end{figure}

The next correlation function which can be used to extract
information on the bound state is the EMDF. Figures~\ref{nq2}(a),(b)
show the EMDF at $J/t=0.3$ and $J/t=1.0$ in the first quadrant of the
Brillouin zone. 
Since the total momentum of the system ${\bf P}$ is zero, the EMDF
possesses the full square symmetry. Moreover, since the ground state
is a singlet, 
$\langle n_{{\bf k}\uparrow}\rangle=
\langle n_{{\bf k}\downarrow}\rangle=
\langle n_{\bf k}\rangle$.
Another noticeable difference from
the single-hole EMDF is the absence of  ``dip''  at any
${\bf k}$-point. This is not surprising because one would not
expect the holes in the bound state to have a certain
momentum. They will be spread over all ${\bf
k}$-points especially if the bound state is well localized in
real space.

Some of the features of the EMDF are essentially the same as that of the
single-hole case. The dome structure is very pronounced. 
Further, our results shows that the
amplitude of the background deviation,
$\Delta n= (\langle n_{(0,0)}\rangle -\langle
n_{(\pi,\pi)}\rangle)$, is roughly the same as
$(\Delta n^{1hole}_{\uparrow}+\Delta n^{1hole}_{\downarrow})$.
This shows that the background behaviour is due to the single-hole
excitations and is irrelevant to the physics of the bound
state. We will provide a support to this in the next Sections.

In the next section we will show that the important 
EMDF data are those along the AF Brillouin zone boundary. These data
are practically unaffected by the kinematic form factor effect, so
they can be used to draw conclusions on the internal structure of the bound
state in ${\bf k}$-space. One can interpret the EMDF at these 
points as the half-filled EMDF suppressed by the 
hole-occupation number. 
The hole weight at the single-hole ground state momentum $(\pi/2,\pi/2)$
is surprisingly small ---
$\langle n_{\bf k}\rangle$ deviates from the half-filled value
of $\frac{1}{2}$ by only 1\%. This is the consequence of the
$d_{x^2-y^2}$ symmetry which restrict the hole weight to be zero at
these points. Another interesting feature is
that the hole occupation at the $(3\pi/4,\pi/4)$ point is higher than
that at  $(\pi,0)$  (Fig. \ref{nq2}(a)). It is worth noting
that at smaller $J/t$ the hole occupation numbers at these points
are very similar and their absolute values are larger. 
As it follows from the discussion in the next sections, these
facts indicate the presence and importance of higher harmonics
in the bound state at smaller $J/t$, because the
``bare'' first $d$-wave harmonic $(\cos(k_x)-\cos(k_y))$ will always give 
a larger hole weight at $(\pi,0)$ than at $(3\pi/4,\pi/4)$.

\begin{figure}
\unitlength1cm
\epsfxsize=6cm
\begin{picture}(8,9)
\put(0.5,0){\epsffile{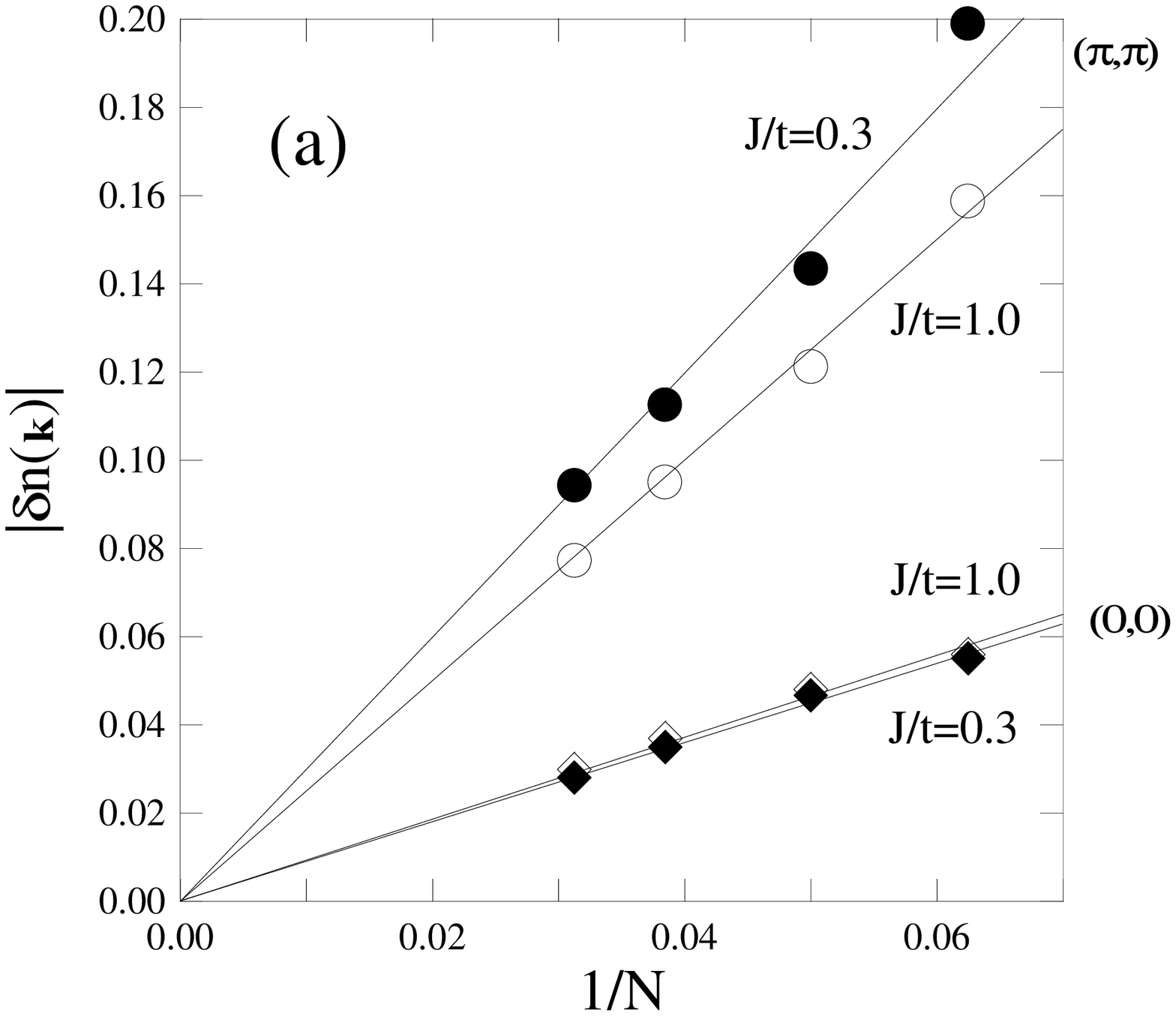}}
\end{picture}
\unitlength1cm
\epsfxsize=6cm
\begin{picture}(8,9)
\put(2,0){\epsffile{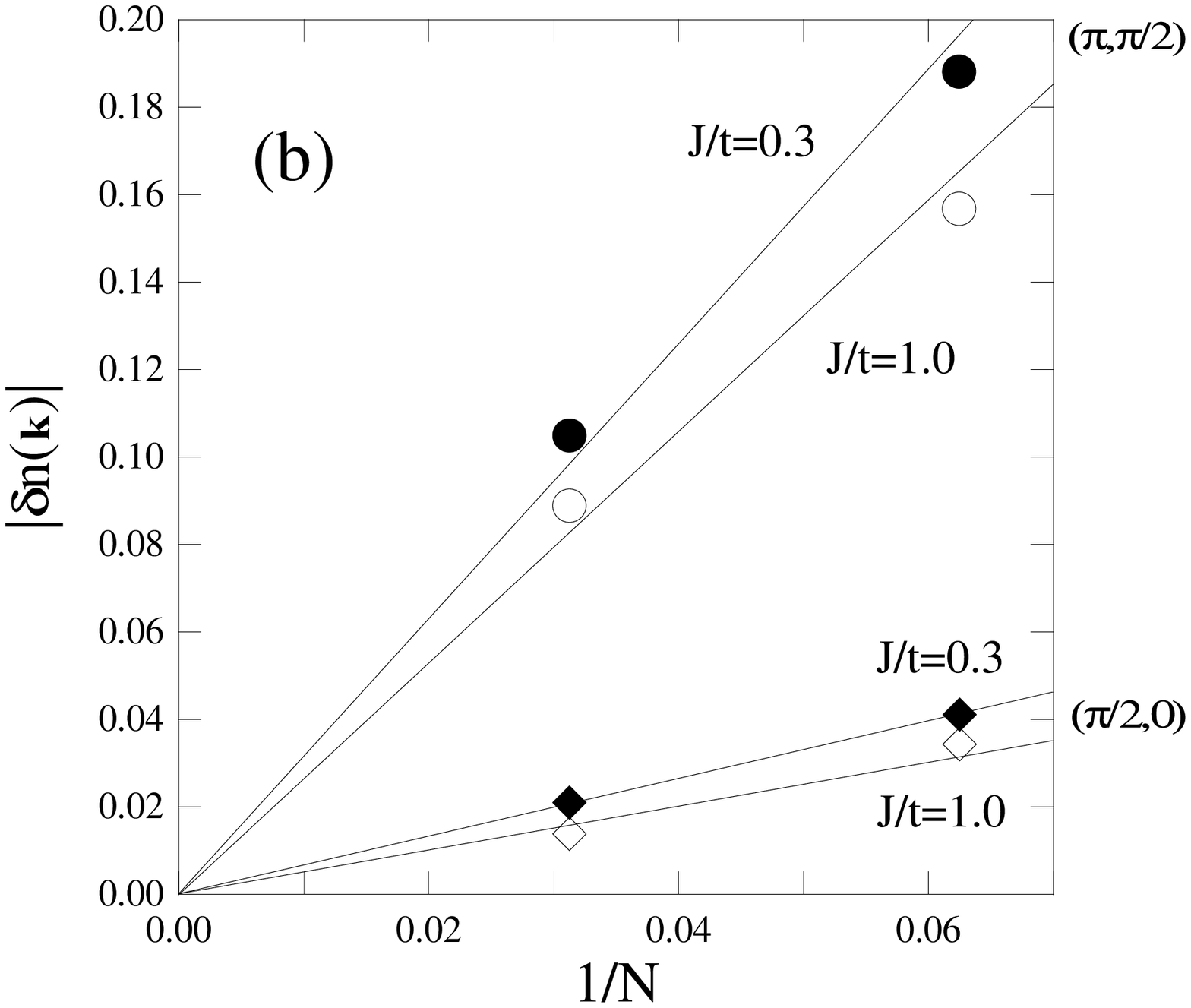}}
\end{picture}
\caption{$1/N$ scaling of the  
$|\langle\delta n_{\bf k}\rangle|$ in the two-hole ground state, 
for (a) ${\bf k}=(\pi,\pi)$ at $J/t=0.3$ (filled circles) and $J/t=1.0$
(empty circles), and $(0,0)$ at $J/t=0.3$ (filled diamonds) and $J/t=1.0$
(empty diamonds), and (b) ${\bf k}=(\pi,\pi/2)$ at $J/t=0.3$ (filled
circles) and $J/t=1.0$ (empty circles), and $(\pi/2,0)$ at $J/t=0.3$ 
(filled diamonds) and $J/t=1.0$ (empty diamonds).}
\label{n2ab}
\end{figure}

\begin{figure}
\unitlength1cm
\epsfxsize=6cm
\begin{picture}(8,9)
\put(5,0.5){\epsffile{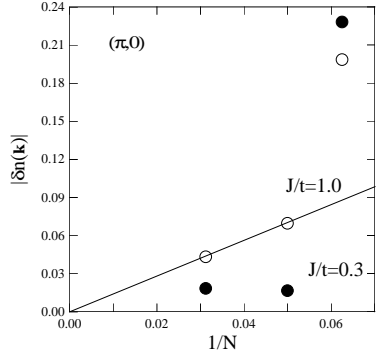}}
\end{picture}
\caption{$|\langle\delta n_{\bf k}\rangle|$
 in the two-hole ground state {\it vs.} $1/N$
for ${\bf k}=(\pi,0)$ at $J/t=0.3$ (filled circles) and $J/t=1.0$
(empty circles). Solid line shows $1/N$ scaling for the $J/t=1.0$
data if the 16-site cluster result is ignored.}     
\label{nq2_pi0}
\end{figure}

The available clusters allow us to perform FSS for
six of the nine inequivalent ${\bf k}$-points of the 32-site cluster.
Results for four of them at  $J/t=0.3$ and $J/t=1.0$ are
presented in Figs. \ref{n2ab}(a),(b). They all show the anticipated $1/N$
scaling. Note that a similar scaling plot at $(\pi/2,\pi/2)$
is not successful because $|\langle\delta n_{\bf k}\rangle|$
is too small. Figure~\ref{nq2_pi0} shows the scaling of the EMDF at
$(\pi,0)$. If we discard the 16-site data by arguing that they
are spoiled by the artificial degeneracy, one can clearly see the 
$1/N$ scaling at $J/t=1.0$. In contrast to this, $|\langle\delta
n_{(\pi,0)}\rangle|$ at $J/t=0.3$ does not show the same the $1/N$ scaling.
We attribute these different
behaviours to the different sizes of the bound states.
The $J/t=1.0$ bound state is small. Therefore, it has to
scale as $1/N$ even when $N$ is not too large. The
$J/t=0.3$ bound state is relatively large. An increase in the cluster
size redistributes the hole weight among the new harmonics which
become available on larger systems. The EMDF at those points
not along the AFBZ boundary (Figs. \ref{n2ab}(a),(b)) 
mostly result from kinematic effects which are saturated at  shorter
distances. Therefore, they do not depend much on the details of the bound
state structure.  
\newpage
\section{Analytical results}

Studies of the doped $t-J$ model via the ED technique
provide important information on effective quasiparticle theories.
However, these same numerical results also posed some problems and made 
questionable the relation of these
analytical studies to the problem of the ``finite
doping of the finite system''. 
For example, the EMDF for the 
ground states of the different
number of holes and the pair correlation function for the
two holes doped into the system were intensively studied numerically
(see Secs. II.A, III). 
It turned out that the results for these quantities were found to be
in the contradiction with some expectations. 
 EMDF, which was naively expected to  show
something like ``hole pockets'', or simply  hole-rich and
electron-rich regions in ${\bf k}$-space, demonstrates  a
dramatic deviation of this quantity for the doped clusters from the
half-filled (no holes) case with the strong variation across the whole
Brillouin zone. Moreover, there is a strong doping dependence of these
results. Data for the two-hole ground state differ
significantly from the single-hole ones. More surprisingly, the overall
shape of the EMDF reminds one of the free electrons with a
nearest-neighbour hopping band. This was the reason for the
conjecture that the $t-J$ model already at rather low doping
concentration undergoes a transition to the free-electron physics
and shows a ``large'' Fermi surface \cite{SH}. 
Also, the 
hole-hole correlation function for the $d$-wave bound state show the
largest weight of the holes in the configurations which should be
forbidden by the $d$-wave symmetry (the so called
``$\sqrt{2}$-paradox'').

In this situation, a physical explanation of such puzzling behaviour
of the correlation functions together with an analytical picture would
be highly desirable.

A qualitative understanding of these effects in the context of the
spin-polaron physics 
 has been achieved in the works by Eder and Becker
\cite{EB}, and Eder and Wrobel \cite{EW}, wherein the authors showed
that the $t-J$ model quasiparticles will favour qualitatively the same
EMDF as the ones found in the numerical calculations. Using rather general
arguments, they demonstrated that the ``large
Fermi-surface'' is a consequence of simple sum rules and a minimum
of the total energy, and it is completely irrelevant to the problem of the
real Fermi-surface identification. 
The main idea of these works is that the hole-pockets
should be attributed to the quasiparticles, not to the
``bare'' holes. Since the renormalization is strong only a relatively
small part of the polaron can be visualized in ${\bf k}$-space as a fermion
having the certain momentum. The ``dressed'' part of the spin-polaron
is responsible for the background in the EMDF, which is spread over
the entire BZ. More specifically, the EMDF does not only measure the
lack of the electrons due to the centre of the polaron, but it
also keeps track of the hole distribution inside the
polaron. Similar physics has been discussed recently in Ref. \cite{RD}.

Within the same theoretical framework, i.e. a variational string
approach, we mention that 
the pairing problem for two holes has been considered elsewhere
\cite{Ed} and the source of the large probability of finding holes in the
ground state along the diagonal of an elementary square can
be explained by the large weight of the ``hole$\times$(hole+1
spin flip)'' combination in the two-hole $d$-wave bound state wave
function. Qualitative discussion of the same physics has been done
recently in Ref. \cite{RD1}.

In what follows we will show how the qualitative picture drawn in
Refs. \cite{EB,EW,Ed}, which gives a basic understanding of the
numerical data, can be reproduced using simple ansatzes for the spin
polarons and their bound state. Then, the CT approach is used to
derive analytical expressions which are able to explain {\it
  quantitatively} most of the 1- and 2-hole ED data for the ground
states described in Sec. II, and earlier in the literature.

\subsection{Qualitative analysis using a simplified model:}

We begin our analytical calculations of two holes described
by the $t-J$ model by considering a simplified treatment of
two holes moving in an AFM Ising background. We then
evaluate
the EMDF and $C(r)$ using this simplified model in order to see what
kind of behaviour one can expect in the ground
state of spin polarons. This work
is instructive, and helps in understanding this problem.
So we present these preliminary results first.

Consider the EMDF. In a system without holes, one finds that
$\langle n_{{\bf k}\sigma}\rangle = 1/2$ everywhere in the full
Brillouin zone.
This is a consequence of the purely local character of the 
electronic states and $S_{tot}^z=0$. By definition  
$\langle n_{{\bf k}\sigma}\rangle=\frac{1}{N}\sum_{ij}e^{i{\bf k}{\bf r}_{ij}}\langle 
\tilde{c}_{i\sigma}^{\dag}\tilde{c}_{j\sigma} \rangle 
=\frac{1}{N}\sum_{i}\langle\tilde{c}_{i\sigma}^{\dag}
\tilde{c}_{i\sigma} \rangle +
\frac{1}{N}\sum_{i,d\not= 0}e^{i{\bf k}\cdot{\bf d}}\langle 
 \tilde{c}_{i\sigma}^{\dag}\tilde{c}_{i+d\sigma} 
\rangle$. 
The second term is zero for the half-filled case and the
first term yields 
$\langle n_{{\bf k}\uparrow}\rangle=\langle n_{{\bf k}\downarrow}\rangle=1/2$. 
An informative result which follows from the second term is that 
hole doping makes the matrix elements between different 
``strings'' of the polaron wave function nonzero, and 
accompanied by the phase factor
$e^{i{\bf k}\cdot{\bf d}}$, where $|{\bf d}|$ is the difference between
the lengths of the strings. For example, the
matrix element between the
bare component 
$\left(\sum_i \tilde{c}_{i\downarrow}e^{i{\bf k}{\bf r}_i}
|0\rangle\right)$ 
and the one spin-flip string component 
$\left(\sum_{i,\delta}S_i^+ \tilde{c}_{i+\delta\uparrow}
e^{i{\bf k}{\bf r}_i}|0\rangle\right)$
is proportional to $\sum_{\delta}e^{i{\bf
k}\mbox{\boldmath $\delta$}}
\sim \gamma_{\bf k}$, which is asymmetric with respect to the transformation
${\bf k}\rightarrow {\bf k}+(\pi,\pi)$,
$\gamma_{\bf k}=-\gamma_{\bf k+Q}$. Thus all
odd-distance matrix elements are responsible for the
antisymmetric contribution to $\langle n_{\bf k}\rangle$, 
and this asymmetry makes $\langle n_{\bf k}\rangle$
resemble the shape of a large Fermi surface. Note, that this
unusual effect is closely related to the localized character of the
electronic states and the spin polaron nature of the carriers. 
Recently, a similar asymmetry observed in the ARPES data of an
undoped (Sr$_2$CuO$_2$Cl$_2$) \cite{SSE} and doped \cite{RD} AFM
insulators has been successfully explained using essentially
the same ideas.

Hole excitations near half-filling (when 
long-ranged AFM order is present) are most concisely explained using the 
spinless-hole, Schwinger-boson representation for the constrained fermion 
operators. Thus it is necessary
to express $\langle n_{\bf k}\rangle$ and $C(r)$ in terms of averages of 
combinations of the hole and magnon operators.
The essence of this representation is the following. The creation of
a hole (annihilation of an electron) at site $i$ in
sublattice $A=\{\uparrow\}$ (with the main direction of the spins being up)
is achieved by operating $\tilde{c}_{i\uparrow}$ on the ground
state. Thus, $\tilde{c}_{i\uparrow}\simeq h^{\dag}_i$. The action of the
same operator on site $j$ in sublattice $B=\{\downarrow\}$ is non-zero only if
the spin is in the ``wrong'' direction ($\uparrow$). Therefore,
creation of a hole is accompanied by the annihilation of a spin
excitation: $\tilde{c}_{j\uparrow}=h^{\dag}_j S^-_j\simeq h^{\dag}_j
a_j$. Thus, 
\begin{eqnarray}
\label{3_0}
\langle\tilde{c}_{i\uparrow}^{\dag}\tilde{c}_{j\uparrow} 
\rangle=&&\langle h_{A,i}h_{A,j}^{\dag}(1-a_{A,j}^{\dag}a_{A,j})
\rangle\delta_{i,A}\delta_{j,A}+
\langle h_{A,i}h_{B,j}^{\dag}a_{B,j}\rangle\delta_{i,A}\delta_{j,B}
\nonumber\\
&&+
\langle h_{B,i}a_{B,j}^{\dag}h_{A,j}^{\dag}\rangle\delta_{i,B}\delta_{j,A}+
\langle h_{B,i}h_{B,j}^{\dag}a_{B,j}^{\dag}a_{B,j}
\rangle\delta_{i,B}\delta_{j,B}
\end{eqnarray}
The above will be suffice for the description in this paper --- for 
an advanced and detailed discussion of this representation 
we refer the readers to Ref. \cite{Onuf}.

First we examine a simple ansatz for the single-hole ground state
wave function \cite{Sush1,SD} 
\begin{eqnarray}
\label{3p}
|1\rangle=\sqrt{\frac{2}{N}}\tilde{h}_{B,{\bf P}}^{\dag}|0\rangle 
=\sqrt{\frac{2}{N}}\biggl[\alpha h_{B,{\bf P}}^{\dag} + 4\beta\sum_{\bf q}
\gamma_{\bf P-q}h_{A,{\bf P-q}}^{\dag}a_{B,{\bf q}}^{\dag}
\biggr] |0\rangle\ , 
\end{eqnarray}
where $\alpha^2+4\beta^2=1$, and,
as noted in Ref. \cite{EB}, the sign of the term linear in 
$\gamma_{\bf k}$ is found from minimizing the kinetic energy. 
(Note that the origin of the hole is 
in sublattice $B$, so the total spin of the system is
$S_{tot}^z=1/2$.)
Hereafter, ${\bf q} \in ABZ$
\begin{eqnarray}
\label{3_1}
\sum_{\bf q}=\frac{2}{N}\sum_{q_n, n=1}^{N/2},\ \ 
{\rm and} \ \ \{h_{A(B){\bf k}},h_{A(B){\bf k}^{\prime}}^{\dag}\}=
[a_{A(B){\bf k}},a_{A(B){\bf k}^{\prime}}^{\dag}]=
\delta_{{\bf k},{\bf k}^{\prime}} \cdot N/2 \ .
\end{eqnarray}
Minimal algebra for the EMDF and $C(r)$ yields
\begin{eqnarray}
\label{3s}
&&\langle n_{{\bf k}\uparrow}\rangle N\simeq 
\frac{N}{2}-\langle h_{A,{\bf k}}^{\dag}
h_{A,{\bf k}}\rangle+\sum_{\bf q}\bigl(\langle a_{B,{\bf q}}^{\dag}
a_{B,{\bf q}}\rangle-\langle a_{A,{\bf q}}^{\dag}
a_{A,{\bf q}}\rangle \bigr)-
\bigl(\langle h_{A,{\bf k}}^{\dag}\sum_{\bf q}
h_{B,{\bf k+q}}a_{B,{\bf q}}^{\dag}\rangle
+\mbox{H.c.} \bigr) ,\nonumber\\
&&\langle n_{{\bf k}\downarrow}\rangle N
\simeq \frac{N}{2}-\langle h_{B,{\bf k}}^{\dag}
h_{B,{\bf k}}\rangle+\sum_{\bf q}\bigl(\langle a_{A,{\bf q}}^{\dag}
a_{A,{\bf q}}\rangle-\langle a_{B,{\bf q}}^{\dag}
a_{B,{\bf q}}\rangle \bigr)-\bigl(
\langle h_{B,{\bf k}}^{\dag}
\sum_{\bf q} h_{A,{\bf k+q}}a_{A,{\bf q}}^{\dag}\rangle
+\mbox{H.c.}\bigr) ,
\nonumber\\
&&C(r)=\frac{1}{N_hN_E(r)}\sum_{i,j}
\langle n^h_i n^h_j\delta_{|i-j|,r}\rangle,
\end{eqnarray}
where $n^h_i=h^{\dag}_i h_i$ is the hole number operator.
The physical meanings of the terms
 in $\langle n_{\bf k}\rangle$ are apparent.
The number of electrons with spin up and momentum ${\bf k}$ is
reduced by the amount of holes having the same
momentum and by the spin flips in 
sublattice $A$. It is increased by the number of spin excitations in
sublattice $B$. The last term is not zero between different
components of the spin-polaron wave function, reflecting the inner
structure of this quasiparticle, or the kinematic ``form-factor''.
Alternatively, according to Ref. \cite{EB}, 
it reflects ``the fast movement of the 
hole inside the bag''. 

Using $|1\rangle$ from Eq.~(\ref{3p}), 
Eq.~(\ref{3s})  give
\begin{eqnarray}
\label{3q}
&&\langle n_{{\bf k}\uparrow}\rangle\simeq \frac{1}{2}+\frac{1}{N}\left(
-16\beta^2\gamma_{\bf k}^2 + 4\beta^2+8|\alpha\beta |\gamma_{\bf
k}\right) ,\\
&&\langle n_{{\bf k}\downarrow}\rangle\simeq \frac{1}{2}-\frac{1}{2}\alpha^2 
\delta_{{\bf k},{\bf P}}- \frac{1}{N}4\beta^2 .
\nonumber
\end{eqnarray}
These simple expressions already contain significant qualitative
information about the EMDF for the single-hole ground state. There
is a ``dip'' in $\langle n_{{\bf k}\downarrow}\rangle$ at ${\bf
k}={\bf P}$ with weight equals to one-half of the quasiparticle
residue, corresponding to the centre of the polaron. There is also
a constant positive (negative) shift in $\langle n_{{\bf k}\uparrow}\rangle$
($\langle n_{{\bf k}\downarrow}\rangle$) due to spin excitations in
sublattice $B$ (see Eq.~(\ref{3p})). 
Although $\langle n_{{\bf k}\uparrow}\rangle$ does 
not have any ``dips'', it does have two other features. One is due 
to the hole distribution in the ``dressed'' part of the polaron 
($\sim\gamma_{\bf k}^2$), and the other is due to the ``interstring'' 
matrix elements ($\sim\gamma_{\bf k}$).   The absence of the
``interstring'' terms in  
$\langle n_{{\bf k}\downarrow}\rangle$ in Eq.~(\ref{3q}) is due to the approximation made 
in the above ansatz {\it viz.} Eq.~(\ref{3p}), namely the elimination
of the longer strings which are 
necessary to produce the ``dome'' structure of 
$\langle n_{{\bf k}\downarrow}\rangle$. 
This explains the smaller amplitude and stronger
$t/J$ dependence of the difference
$\Delta n_{\downarrow}=n_{(0,0)\downarrow}-n_{(\pi,\pi)\downarrow}$
than those of $\Delta n_{\uparrow}$ for the single-hole GS.

Thus, most features of the single-hole EMDF data reported in Sec. III
can be understood using this simplified model. According to
Eq.~(\ref{3q}) all $\langle \delta n_{{\bf k}\sigma}\rangle$ scale as $1/N$
except for the dip which scales as $C+\alpha/N$, 
a result which is employed in the FSS analysis of the ED numerical work.

In order to carry out similar analysis for the  two hole case, 
one has to solve Schr\"{o}dinger equation for the bound state problem. 
Instead of doing this we simply propose the nearest-neighbour bound
state wave function having $d$-wave symmetry and $S^z_{tot}=0$
based on the expectation that two static holes attract each other 
through the ``sharing common link'' 
effect ({\it viz.}, $H_{int} = - J/2~n^h_i n^h_j$): 
\begin{eqnarray}
\label{5w}
|2\rangle =&&\sqrt{\frac{2}{N}}
\sum_{\bf p}\Delta^{d}_{\bf p}\tilde{h}_{A,{\bf p}}^{\dag}
\tilde{h}_{B,{\bf -p}}^{\dag} |0\rangle
\nonumber\\
&&=\sqrt{\frac{2}{N}}\sum_{\bf p}\Delta^{d}_{\bf p}
\biggl[\alpha^2 h_{A,{\bf p}}^{\dag} h_{B,{\bf -p}}^{\dag} 
+ 4\alpha\beta\sum_{\bf q}\left(\gamma_{\bf p-q}
h_{B,{\bf p-q}}^{\dag}h_{B,{\bf -p}}^{\dag}a_{A,{\bf q}}^{\dag}+
\gamma_{\bf -p-q}h_{A,{\bf p}}^{\dag}h_{A,{\bf -p-q}}^{\dag}
a_{B,{\bf q}}^{\dag}\right)\\
&&\phantom{=\frac{1}{\sqrt{N/2}}\sum_{\bf p}\Delta^{d}_{\bf p}\biggl[}
+16\beta^2\sum_{\bf q,q^{\prime}}
\gamma_{\bf p-q}\gamma_{\bf -p-q^{\prime}}h_{B,{\bf p-q}}^{\dag}
h_{A,{\bf -p-q}}^{\dag}a_{A,{\bf q}}^{\dag}
a_{B,{\bf q^{\prime}}}^{\dag}
\biggr] |0\rangle\ ,
\nonumber
\end{eqnarray}
with $\Delta^{d}_{\bf p}=\left(\cos(p_x)-\cos(p_y)\right)$ ensuring
the centres of the polarons are at the nearest-neighbour sites.

The EMDF calculation using $|2\rangle$ in Eq. (\ref{5w}) yields
\begin{eqnarray}
\label{3w}
\langle n_{\bf k}\rangle=\langle n_{{\bf k}\uparrow}\rangle=\langle n_{{\bf k}\downarrow}\rangle
\simeq \frac{1}{2}+\frac{1}{N}\left(-\alpha^2(\Delta^{d}_{\bf k})^2
-16\beta^2\gamma_{\bf k}^2 + 8|\alpha\beta |\gamma_{\bf k}\right) ,
\end{eqnarray}
where the terms inside the bracket are simply the sum of the $1/N$ terms
in the $\langle n_{{\bf k}\uparrow}\rangle$ and $\langle n_{{\bf k}\downarrow}\rangle$ expressions
of Eq.~(\ref{3q}) for the single-hole case, and the ``dip'' structure 
is replaced by the probability of finding a ``bare'' hole with momentum 
${\bf k}$ in the bound state. This explains the observation mentioned
in Sec. III that the quantity 
$\Delta n=n_{(0,0)}-n_{(\pi,\pi)}$ for the two-hole case 
is roughly the same as
$\Delta n_{\uparrow}+\Delta n_{\downarrow}$ for the single-hole case. 
It is interesting to note that since the ${\bf k}$-dependence of the terms 
from the dressed part of the polaron and from the ``interstring'' processes
vanish at the boundary of the magnetic Brillouin zone 
($\gamma_{\bf k}=0$), features of the EMDF
along this line are not disguised by kinematic effects.
Thus, one can directly observe the structure of the bound state wave
function $\Delta^{d}_{\bf k}$ from the
$\langle n_{\bf k}\rangle$ data at these points. 
In particular, the $(\pi/2,\pi/2)$ point
has to have zero hole weight due to the $d$-wave symmetry
of the bound state. For the particular form of $\Delta^{d}_{\bf p}$ we
have chosen, the maximum of the hole weight (minimum in $\langle n_{\bf
  k}\rangle$) will be at $(\pi,0)$ point.

The hole-hole correlation function on different clusters consistently
shows a maximal probability for states in which the holes
are along the diagonal of an elementary square, i.e., they
prefer to be at a distance ${\sqrt {2}}a$ from one another, where $a$ is the 
lattice constant.
(At first glance, such a configuration should be forbidden
by the $d_{x^2-y^2}$-symmetry of the state. 
One way to resolve this paradox, as suggested by Poilblanc 
\cite{p94},
is to introduce
modified creation pair operators
$h_i^{\dag}h_{i\pm x\pm y}^{\dag}S_{i\pm x(y)}^+$ to the ``bare'' 
$h_i^{\dag}h_{i\pm x(y)}^{\dag}$ pair operator. It is clear that the
bound state wave function Eq. (\ref{5w}) includes such combinations
naturally.) Calculation of $C(r)$ of Eq.~(\ref{3s}) in the ground state 
given by Eq.~(\ref{5w}) gives 
\begin{eqnarray}
\label{Cr}
C(1)=\alpha^4/4 + 9\beta^4/4 , \ C(\sqrt{2})=\alpha^2\beta^2, \
C(2)=\alpha^2\beta^2/2, \ C(\sqrt{5})=3\beta^4/4 ,\ 
C(3)=\beta^4/4 .
\end{eqnarray}
For the physical range of $t/J\sim 2-3$ the weights 
accumulated in the ``bare'' and ``1-string'' parts of the polaron wave
function are almost identical, $\alpha^2\simeq 4\beta^2$ \cite{Sush1,FKS}.
This gives $C(1)\simeq C(\sqrt{2})$,   in
qualitative agreement with the numerical results. 

Thus, one can conclude that our simple considerations 
of one and two holes in a system of Ising
spins, based on a simplified spin-polaron picture, already shows
qualitative agreement with the numerical data. 
The treatment of the realistic system with a N\'{e}el spin
background requires a proper account of the spin fluctuations, 
the long-range dynamics of the system, and multiple spin excitations 
(longer ``strings'').

\subsection{CT approach}
\subsubsection{CT Hamiltonian:}

The $t-J$ model Hamiltonian (\ref{hamiltonian}) can be rewritten using the
spinless-fermion representation for the constrained fermion operators
and Holstein-Primakoff \cite{KLR,Horsch} or Dyson-Maleev \cite{Onuf} 
representation for the spin operators. These formalisms have been shown 
to be adequate in treating the nonlinear feature of the kinetic 
energy term of Eq. (\ref{hamiltonian}) properly. 
Subsequent diagonalization of the 
spin part of the Hamiltonian by the Bogoliubov transformation  
naturally includes spin-fluctuations in the ground state.

The essential part of the $t-J$ Hamiltonian rewritten in this way
looks like the electron-phonon Hamiltonian for the ``usual'' polaron
problem with an additional direct fermion-fermion interaction
term:
\begin{eqnarray}
\label{1a}
&&{\cal H}_{t-J}\simeq \sum_{\bf q}\omega_{\bf q} \alpha_{\bf
q}^{\dag}\alpha_{\bf q} + \sum_{{\bf k},{\bf q}}\left(M_{{\bf 
k},{\bf q}} h_{\bf k-q}^{\dag} h_{\bf k}\alpha_{\bf q}^{\dag} + 
\mbox{H.c.}\right)+\Delta H \ ,\\
&&\Delta H= -2J(1-2\lambda)\sum_{{\bf k},{\bf k^{\prime}},{\bf q}}
\gamma_{\bf q} h_{\bf k-q}^{\dag}h_{\bf k^{\prime}+q}^{\dag} h_{\bf
  k^{\prime}} h_{\bf k} \ , \nonumber
\end{eqnarray}
where $h^{\dag} (h)$, $\alpha^{\dag} (\alpha)$, are the spinless 
hole and magnon operators, respectively,
$\omega_{\bf q}=2J(1-\gamma_{\bf q}^2)^{1/2}$ is the spin-wave energy, 
$M_{{\bf k},{\bf q}} = 4 t (\gamma_{\bf k-q}U_{\bf q}+\gamma_{\bf k} 
V_{\bf q})$, $U_{\bf q}, V_{\bf q}$ are the Bogoliubov   
transformation parameters, $\gamma_{\bf k}=(\cos k_x+\cos k_y)/2$,
$\lambda=\sum_{\bf q}(V_{\bf q}^2-\gamma_{\bf q}U_{\bf q}V_{\bf q})=-0.08$.
$\Delta H$ is an effective hole-hole attraction due to minimization
of the number of broken AF bonds. 
Two important differences make the $t-J$ version of the polaron
problem much more difficult to study: (i) the absence of ``bare''
dispersion term of the hole \cite{YuLu}, and (ii) the essentially 
non-local character of the hole-magnon interaction, because each
process of emitting (absorbing) a magnon is associated with an
intersite hole hopping. 

The CT approach has been applied to the spin-polaron Hamiltonian 
of Eq.~(\ref{1a}) in Ref. \cite{bel97}.
The generator of the CT  was proposed to be in the form
\begin{eqnarray}
\label{4C}
 {\cal S}= \sum_{{\bf k},{\bf q}} f_{\bf k}M_{{\bf k},{\bf 
q}}\left(h_{\bf k-q}^{\dag} h_{\bf k} \alpha_{\bf q}^{\dag} 
- \mbox{H.c.}\right),
\end{eqnarray}
where $f_{\bf k}$ is the parameter of the transformation,
and in Ref. \cite{bel97} $f_{\bf k}$ was chosen to minimize the single
hole energy, {\it viz.}
\begin{eqnarray}
\label{N7}
\frac{\delta }{\delta f_{\bf k}}\left(
 \sum_{\bf k^{\prime}} E_{\bf k^{\prime}}\right)
= 0 \ .
\end{eqnarray}
The negligible role of the higher order hole-magnon vertices in the
transformed Hamiltonian
was demonstrated and it was argued that the initially strong
hole-magnon interaction in Eq. (\ref{1a}) is transfered mainly 
into a hole "dressing" and into the hole-hole 
interaction. Thus, for a wide region of $t/J$ one can restrict one's
considerations to the effective Hamiltonian 
\begin{eqnarray}
\label{4a}
{\cal H}_{eff}=\sum_{\bf k}E_{\bf k}
\tilde{h}_{\bf k}^{\dag} \tilde{h}_{\bf k}+ \sum_{\bf q} \omega_{\bf q}
\alpha_{\bf q}^{\dag} \alpha_{\bf q}
+ \sum_{{\bf k},{\bf k^{\prime}},{\bf q}}
V_{{\bf k},{\bf k^{\prime}},{\bf q}}
\tilde{h}_{\bf k-q}^{\dag} 
\tilde{h}_{\bf k^{\prime}+q}^{\dag} \tilde{h}_{\bf k^{\prime}} 
\tilde{h}_{\bf k}
+  \sum_{{\bf k},{\bf q}}F_{\bf {\bf k},{\bf q}}
M_{{\bf k},{\bf q}}\left(\tilde{h}_{\bf k-q}^{\dag} \tilde{h}_{\bf k}
\alpha_{\bf q}^{\dag} + \mbox{H.c.}\right)
\end{eqnarray}
where $E_{\bf k}$ and $\omega_{\bf q}$ are the polaron and magnon energies
respectively, $V_{{\bf k},{\bf k^{\prime}},{\bf q}}$ is the direct
polaron-polaron interaction, $M_{{\bf k},{\bf q}}$ is the bare hole-magnon 
vertex, and $F_{\bf {\bf k},{\bf q}}$ is the renormalization form-factor which 
is close to zero at large ${\bf q}$, and is constant ($\sim 0.2-0.4$) at 
small ${\bf q}$. The last term, which corresponds to the interaction 
of the hole with
long-range spin waves, has been left in the effective Hamiltonian in this 
form  to account for the retardation effect in the polaron-polaron
spin-wave exchange. Also, short-range spin-wave exchange has been converted 
to the direct polaron-polaron interaction. The polaron energy $E_{\bf k}$ and 
the 
weights of the components of the polaron's wave function have been compared 
with the results of the other works, especially SCBA results, and  very good 
agreements were found. Since the derivation of the single polaron energy 
{\it and} the polaron-polaron interaction in the framework of CT
approach are the same, one can hope that the effective 
Hamiltonian of Eq. (\ref{4a}) properly describes the interaction between the 
low-energy excitations of the $t$-$J$ model. 

\subsubsection{Our calculations using the CT approach}

We are interested in the ground state with total spin
$S_{tot}^z=1/2$ ($S_{tot}^z=0$) for the single-hole (two-hole) case
in an AF ordered system. Thus, it is necessary to use a two-sublattice
representation for the fermions and bosons \cite{bel97}.
In the two-sublattice representation there are two types of 
holes and magnons, both defined inside the first magnetic Brillouin
zone, whereas in the one-sublattice representations holes and magnons
are defined inside the full Brillouin zone.
In the previous subsection we used the latter for the sake of
simplifying notations. There is a simple relation between these two
representations:
\begin{eqnarray}
\label{T1}
&&h_{\bf k}=(f_{\bf k}+g_{\bf k})/\sqrt{2}\ , \ \ \
h_{{\bf k}+(\pi,\pi)}=(f_{\bf k}-g_{\bf k})/\sqrt{2}  \ , \\
&&a_{\bf q}=(\alpha_{\bf q}+\beta_{\bf q})/\sqrt{2}\ , \ \ \
a_{{\bf q}+(\pi,\pi)}=(\alpha_{\bf q}-\beta_{\bf q})/\sqrt{2} \ ,
\nonumber
\end{eqnarray}
where $f_{\bf k}$ and $g_{\bf k}$ correspond to the hole excitations in
the $A$ and $B$ sublattices respectively. $\alpha_{\bf q}$ and $\beta_{\bf 
q}$ are the two branches of the Bogoliubov spin-wave excitations.

The correlation functions $\langle n_{{\bf k}\sigma}\rangle$ and $C(r)$ 
expressed in terms of
the averages of the hole and magnon operators are 
\begin{eqnarray}
\label{4s}
&&\langle n_{{\bf k}\uparrow}\rangle N\simeq \frac{N}{2}-(1-\delta\lambda)
\langle f_{\bf k}^{\dag}f_{\bf k}\rangle-\sum_{\bf q}\langle
g_{\bf k+q}^{\dag}g_{\bf k+q}\rangle V_{\bf q}^2+
\sum_{\bf q}\left(\langle \beta_{\bf q}^{\dag}
\beta_{\bf q}\rangle-\langle \alpha_{\bf q}^{\dag}
\alpha_{\bf q}\rangle \right) \nonumber\\
&&\phantom{\langle n_{{\bf k}\uparrow}\rangle\simeq \frac{1}{2}}
-\bigl(\langle f_{\bf k}^{\dag}\sum_{\bf q}
g_{\bf k+q}(\beta_{\bf q}^{\dag}
U_{\bf q}+\alpha_{\bf -q}V_{\bf q})\rangle
+\mbox{H.c.}\bigr) ,\nonumber\\
&&\langle n_{{\bf k}\downarrow}\rangle N\simeq \frac{N}{2}-(1-\delta\lambda)
\langle g_{\bf k}^{\dag}g_{\bf k}\rangle-\sum_{\bf q}\langle
f_{\bf k+q}^{\dag}f_{\bf k+q}\rangle V_{\bf q}^2+
\sum_{\bf q}\left(\langle \alpha_{\bf q}^{\dag}
\alpha_{\bf q}\rangle-\langle \beta_{\bf q}^{\dag}
\beta_{\bf q}\rangle \right) \nonumber\\
&&\phantom{\langle n_{{\bf k}\downarrow}\rangle\simeq \frac{1}{2}}
-\bigl(\langle g_{\bf k}^{\dag}\sum_{\bf q}
f_{\bf k+q}(\alpha_{\bf q}^{\dag}
U_{\bf q}+\beta_{\bf -q}V_{\bf q})\rangle
+\mbox{H.c.}\bigr) ,\nonumber\\
&&C(r)=\frac{1}{N_hN_E(r)}\sum_{i,j}
\langle n^h_i n^h_j\delta_{|i-j|,r}\rangle,
~~~~h = f(g),\ {\rm when}\ \  i \in \{A\} (\{B\}) ,
\end{eqnarray}
where $\delta\lambda=\sum_{\bf q}V_{\bf q}^2=0.19$.
The negligible contribution of the higher order terms (in the number of
magnons) to $\langle n_{{\bf k},\sigma}\rangle$ 
has been checked and these terms are omitted.  

It is interesting to compare these expressions with those for the Ising
limit of the model given in Eq. (\ref{3q}). The number of holes in
sublattice $A$ reducing the number of electrons with spin up is decreased by 
the spin fluctuations ($\delta\lambda$, first term), but due to the same effect
the reduction in $\langle n_{{\bf k}\uparrow}\rangle$ can be done by the holes in the
sublattice $B$ (second term). The third terms take into account an
imbalance of the number of  spin excitations of  different
types. The last term is nonzero for the ``interstring'' processes
(now ``strings'' are just the components of the wave function with the
spin excitations).

Using the CT generator of Eq. (\ref{4C}) one  obtains
the wave function of the spin polaron ($S_{tot}^z=+1/2$)
\begin{eqnarray}
\label{3a}
|1\rangle=\sqrt{\frac{2}{N}}\tilde{g}_{\bf P}^{\dag}|0\rangle 
=\sqrt{\frac{2}{N}}e^{\cal S}g_{\bf P}^{\dag}|0\rangle
=\sqrt{\frac{2}{N}}\biggl[a_{\bf P}g_{\bf P}^{\dag} + \sum_{\bf q}b_{\bf 
P, q} f_{\bf P-q}^{\dag}\beta_{\bf q}^{\dag}
+\sum_{{\bf q}, {\bf q^{\prime}}}
c_{\bf P, q, q^{\prime}} g_{{\bf P}-{\bf q}-{\bf q^{\prime}}}^{\dag} 
\beta_{\bf q}^{\dag}\alpha_{{\bf q^{\prime}}}^{\dag} + \dots \biggr] 
|0\rangle\ . 
\end{eqnarray}
Here, $a_{\bf P}^2=Z_{\bf P}<1$ is the 
quasiparticle residue. An explicit expression for the exact spin-polaron
wave function within the SCBA was written in Ref. \cite{GR} in the same form. 
The ground state momenta for the
spin-polaron in the pure $t-J$ model are $\pm(\pm\pi/2,\pi/2)$.

Then, using Eq. (\ref{3a}) the single-hole EMDF is found to be given
\begin{eqnarray}
\label{4q}
&&\langle n_{{\bf k}\uparrow}\rangle\simeq \frac{1}{2}+\frac{1}{N}\biggl(
-(1-\delta\lambda) b_{\bf P,P-k}^2 - a_{\bf P}^2 V_{\bf P-k}^2
-\sum_{{\bf q}, {\bf q^{\prime}}} 
c_{\bf P,q^{\prime},P-k-q-q^{\prime}}^2 V_{\bf q}^2  
\nonumber\\
&&\phantom{\langle n_{{\bf k}\uparrow}\rangle\simeq \frac{1}{2}}
+\sum_{\bf q} b_{\bf P,q}^2 -2\bigl(a_{\bf P}
b_{\bf P,P-k}U_{\bf P-k} 
+b_{\bf P,P-k}\sum_{\bf q}c_{\bf P,P-k,q} V_{\bf q}
\bigr)\biggr) ,\\
&&\langle n_{{\bf k}\downarrow}\rangle\simeq \frac{1}{2}-
\frac{1}{2}\delta_{{\bf k},{\bf P}}(1-\delta\lambda)a_{\bf
P}^2+\frac{1}{N}\biggl(-(1-\delta\lambda)\sum_{\bf q}
c_{\bf P,q,P-k-q}^2 -\sum_{\bf q} b_{\bf P,P-k-q}^2 V_{\bf q}^2 
\nonumber\\
&&\phantom{\langle n_{{\bf k}\downarrow}\rangle\simeq \frac{1}{2}}
-\sum_{\bf q} b_{\bf P,q}^2 -2\sum_{\bf q} b_{\bf P,P-k-q}
c_{\bf P,P-k-q,q}U_{\bf q}\biggr) .
\nonumber
\end{eqnarray}
As we will show below, these expressions give  good 
{\it quantitative} agreements with numerical data. 
As before, $\langle n_{{\bf k}\downarrow}\rangle$ shows a ``dip''
at ${\bf k}={\bf P}$ with a weight proportional to 
of the quasiparticle residue due to the centre of the polaron. 
A constant positive (negative) shift due to  different amount of 
spin excitations (fourth term) is also present in $\langle n_{{\bf
k}\uparrow}\rangle$ 
($\langle n_{{\bf k}\downarrow}\rangle$). The first three terms in 
$\langle n_{{\bf k}\uparrow}\rangle$ and the second and third terms in 
$\langle n_{{\bf k}\downarrow}\rangle$ reflect the hole distribution inside the
polaron, whereas the last two terms in $\langle n_{{\bf k}\uparrow}\rangle$ 
and the last term in $\langle n_{{\bf k}\downarrow}\rangle$ are from the
``interstring'' matrix elements. 
As before, they are odd with respect to the transformation ${\bf k}\rightarrow 
{\bf k}+{\bf Q}$ and lead to the formation of the ``dome'' 
structure in the EMDF. As we noted, the asymmetric term in 
$\langle n_{{\bf k}\downarrow}\rangle$ comes from the matrix element between the second 
and third components of the wave function of Eq. (\ref{3a}). 
We restrict ourselves to the first three components of 
Eq. (\ref{3a}) because for $J/t=0.3$ 
they give about 98\% of the norm of the wave function. (We note
that in the SCBA approach the same approximation gives about 
92\% of the norm \cite{RH}).

Formally, Eqs.~(\ref{4q}) give a $1/N$ scaling
for $\langle \delta n_{{\bf k}\sigma}\rangle$ at every ${\bf k}$-point 
except for the ``dip'' in $\langle \delta n_{{\bf
    k}\downarrow}\rangle$ at ${\bf k}={\bf P}$ . 
It fails at the point ${\bf k}={\bf P}-{\bf Q}_{AF}$ where some of the
terms in Eq. (\ref{4q}) are singular.
The reason for these singularities is a peculiarity of the spin-polaron 
ground state and the AF long-range order. The dressing of the hole in the
N\'{e}el background involves an infinite amount of zero-energy 
${\bf q}={\bf Q}$ spin excitations (whose total contribution to the
hole weight is finite and small due to the diminishing magnon density
of states). Since the EMDF probes the inner structure of the
ground state it is actually measuring this singular probability of the
virtual emission of a zero-energy magnon (${\bf Q}$) by the hole
(${\bf P}$) if ${\bf k}$ is equal to  ${\bf P}-{\bf Q}$ \cite{response}. 
This leads to singularities of different types for  
$\langle n_{{\bf k}\uparrow}\rangle$ and $\langle n_{{\bf
    k}\downarrow}\rangle$: 
\begin{eqnarray}
\label{nnq}
\langle n_{{\bf k}\uparrow}\rangle\sim 
\frac{1}{N}\frac{1}{\omega({\bf k-(P-Q)})} , \ \ \ 
\langle n_{{\bf k}\downarrow}\rangle\sim \frac{1}{N} 
\ln(\omega({\bf k-(P-Q)}) ,
\end{eqnarray}
where $\omega({\bf k})$ is the magnon energy.
For the finite system the magnon spectrum has the finite energy gap 
at ${\bf Q}_{AF}$ which scales as \cite{Neuberger,Hasenfratz}
\begin{eqnarray}
\label{E_N}
\Delta E=J\frac{c^2}{\rho_s}\frac{1}{N}\left(1-
\frac{c}{\rho_s}\frac{3.9}{4\pi}\frac{1}{\sqrt{N}}+\dots\right) ,
\end{eqnarray}
where $c\simeq 1.67$ and $\rho_s\simeq 0.175$ \cite{Sandvik} 
are the spin-wave
velocity and spin stiffness, respectively.
This result gives ``antidips'' reported in Sec. III with the 
following scaling laws: 
\begin{eqnarray}
\label{nnq1}
\langle n_{\uparrow}({\bf P-Q})\rangle
\simeq\frac{1}{2}+\frac{C_{\uparrow}}{N}-B_{0\uparrow}-
\frac{B_{1\uparrow}}{\sqrt{N}}+\frac{B_{2\uparrow}}{N} , 
\ \ \ 
\langle n_{\downarrow}({\bf P-Q})\rangle
\simeq\frac{1}{2}-\frac{C_{\downarrow}}{N}-
\frac{B_{\downarrow}\ln(C_0 N)}{N}  ,
\end{eqnarray}
where all constants are positive. 
$C_{\uparrow}$, $C_{\downarrow}$ are from the ``regular'' part of Eq.
(\ref{4q}). An interesting result shown in Eq.~(\ref{nnq1}) is that the
``antidip'' in $\langle n_{\uparrow}({\bf k})\rangle$ is predicted to
survive in the thermodynamic limit: 
\begin{eqnarray}
\label{nnq2}
\langle n_{\uparrow}(-\pi/2,-\pi/2)\rangle\sim 
Z_{\bf P}\rho_s/c^2\simeq 0.07\cdot Z_{\bf P},
\end{eqnarray}
whereas all other features except
for the dip at ${\bf P}$ in 
$\langle n_{\downarrow}({\bf k})\rangle\sim Z_{\bf P}$ will disappear.
One can see from Eq.~(\ref{nnq1}) that the 
scaling laws of the antidips are quite complicated. For a system as
small as $N=32$, terms of different order in $N$ have similar
amplitudes. For example, 
$B_{1\uparrow}/\sqrt{32}\simeq 0.5 B_{0\uparrow}$. This makes the FSS
for the ``antidips'' complicated, especially when only two of
available clusters (16 and 32) possess this ${\bf k}$-point. An
additional complication comes from the fact that the gap $\Delta E$
Eq.~(\ref{E_N}) is 
calculated for the system without holes and the influence of the
latter on it is not known. In the subsequent calculation
of the EMDF for ``antidips'' we modify the magnon spectrum employed in
Eq.~(\ref{4q}) in a way that it has a gap $\Delta E$ at 
${\bf q}={\bf Q}$ point \cite{gap}.

The two-hole problem has been considered using the 
Hamiltonian of Eq. (\ref{4a})
in Ref. \cite{bel97}. A bound state with  $d_{x^2-y^2}$ symmetry was
found for $0<t/J<5$. 
The wave function of the $d$-wave bound state with  total momentum
${\bf P}=0$ can be written in the terms of creation operators of polarons of Eq. (\ref{3a}),
\begin{eqnarray}
\label{5a}
&&|2\rangle =|\Psi_{{\bf P}=0}^{d}\rangle=
\sqrt{\frac{2}{N}}
\sum_{\bf p}\Delta^{d}_{\bf p}\tilde{f}_{\bf p}^{\dag}
\tilde{g}_{\bf -p}^{\dag} |0\rangle \\
&&\Delta^{d}_{\bf p}=\sum_{n=1}^{\infty}\sum_{m=-\infty}^{\infty} 
C_{2n-1,2m}
\left\{\cos([2n-1]p_x+2m p_y)-\cos(2m p_x+[2n-1]p_y)\right\}
\nonumber\\
&&\phantom{\Delta^{d}_{\bf p}=}=C_{1,0}\left\{\cos(p_x)-\cos(p_y)\right\}+
C_{3,0}\left\{\cos(3p_x)-\cos(3p_y)\right\}\nonumber\\
&&\phantom{\Delta^{d}_{\bf p}=}\phantom{=}
+C_{1,2}\left\{\cos(p_x\pm 2p_y)-\cos(p_y\pm 2p_x)\right\}
+\dots \ , \ \ C_{2n-1,2m}=C_{2n-1,-2m}
\nonumber
\end{eqnarray}
where $\Delta^{d}_{\bf p}$  is the solution of an analog of the 
Schr\"{o}dinger equation for the two-body problem. The form of 
$\Delta^{d}_{\bf p}$ ensures the $d$-wave symmetry of the state and 
that the centres of the polarons are always on  different sublattices,
which in  turn guarantees $S^z=0$.
$\Delta^{d}_{\bf p}$ has a more general form than the simple
``nearest-neighbour $d$-wave'' in the simplified
example of Eq. (\ref{5w}). It has higher harmonics,
which have substantial weight for realistic $t/J$.
In what follows we show that
our comparison leads us to the conclusion that the large higher harmonics
of $\Delta^{d}_{\bf p}$ play an important role in
determining the behaviour of $C(r)$ and $\langle n_{\bf k}\rangle$.

Note that for the representative value $J/t=0.3$
about 42\% of the {\it polarons} in the bound state are located at 
the nearest-neighbour sites and less than 2\% are farther than 7
lattice spaces.  An interesting feature of this distribution is that
the  probability in finding the second polaron at a
certain distance from the first falls off slower along the $x$ and $y$
directions. Thus, the weight of the $(3,0)$
($\cos(3p_x)-\cos(3p_y)$) harmonic of $\Delta^{d}_{\bf p}$ is
rather large (20\%), whereas the weight of (1,2) ($\cos(p_x\pm
2p_y)-\cos(2p_x\pm p_y)$) component is less than 5\%.

Finally, relating the previous forms of the wave function (in terms of
creation operators of holes and magnons) Eq. (\ref{5a}) becomes
\begin{eqnarray}
\label{6a}
&&|2\rangle =\sqrt{\frac{2}{N}}
\sum_{\bf p}\Delta^{d}_{\bf p}e^{\cal S}
f_{\bf p}^{\dag} g_{\bf -p}^{\dag} |0\rangle
=\sqrt{\frac{2}{N}}
\sum_{\bf p}\Delta^{d}_{\bf p}\biggl[\hat{\Delta}_0^{\dag}({\bf p})+
\hat{\Delta}_1^{\dag}({\bf p})+\hat{\Delta}_2^{\dag}({\bf p})
+\dots\biggr]|0\rangle \\
&&\phantom{|2\rangle}
\hat{\Delta}_0^{\dag}({\bf p})=
A_{\bf p}^{(1)}f_{\bf p}^{\dag}g_{\bf -k}^{\dag}
+\sum_{\bf q} A_{\bf p,q}^{(2)}g_{\bf k-q}^{\dag}f_{\bf -k+q}^{\dag}+
\sum_{\bf q,q^{\prime}} 
A_{\bf p,q,q^{\prime}}^{(3)}f_{\bf k-q-q^{\prime}}^{\dag}
g_{\bf -k+q+q^{\prime}}^{\dag}+\dots
\nonumber \\
&&\phantom{|2\rangle}
\hat{\Delta}_1^{\dag}({\bf p})=
\sum_{\bf q}\biggl[ B_{\bf p,q}^{(1)}f_{\bf p}^{\dag}f_{\bf
-p-q}^{\dag}
+\sum_{\bf q^{\prime}} B_{\bf p,q,q^{\prime}}^{(2)}
f_{\bf p-q^{\prime}-q}^{\dag}f_{\bf -p+q^{\prime}}^{\dag}
+\dots\biggr]\beta_{\bf q}^{\dag}
\nonumber \\
&&\phantom{|2\rangle\hat{\Delta}_1^{\dag}({\bf p})=}
+\sum_{\bf q}\biggl[ B_{\bf -p,q}^{(1)}g_{\bf p-q}^{\dag}g_{\bf
-p}^{\dag}
+\sum_{\bf q^{\prime}} B_{\bf -p,q,q^{\prime}}^{(2)}
g_{\bf p+q^{\prime}}^{\dag}g_{\bf -p-q-q^{\prime}}^{\dag}
+\dots\biggr]\alpha_{\bf q}^{\dag}
\nonumber \\
&&\phantom{|2\rangle}
\hat{\Delta}_2^{\dag}({\bf p})=
\sum_{\bf q, q^{\prime}}\biggl[ C_{\bf p,q,q^{\prime}}^{(1)}
f_{\bf p-q-q^{\prime}}^{\dag}g_{\bf -p}^{\dag}+
C_{\bf -p,q^{\prime},q}^{(1)}
f_{\bf p}^{\dag} g_{\bf -p-q-q^{\prime}}^{\dag}+
C_{\bf p,q,q^{\prime}}^{(2)}
g_{\bf p-q}^{\dag} f_{\bf -p-q^{\prime}}^{\dag}+
\dots\biggr]\alpha_{\bf q}^{\dag}\beta_{\bf q^{\prime}}^{\dag}
\nonumber
\end{eqnarray}
where the subscripts $n$ of $\hat{\Delta}$'s indicate the number of magnons
in the corresponding component of the wave function.

Results of the  EMDF and $C(r)$ 
in Eq. (\ref{4s}) for the ground state of Eq. (\ref{6a}) are given in full
detail in the Appendix. In the next section we use these 
expressions to compare this theory with the numerical data discussed 
earlier in this paper.

\section{Comparison of numerical and analytical results}

This section summarizes the comparison of our numerical ED data with the
analytical results obtained from the CT approach. We focus on the EMDF 
for one and two holes, the binding energy, and the hole-hole correlation 
function for two holes.  These provide a representative juxtaposition of 
results obtained from these two techniques, and probe in detail the 
correlations found in the ground states.

\begin{figure}
\unitlength1cm
\epsfxsize=7cm
\begin{picture}(4,9)
\put(3,1.5){\rotate[r]{\epsffile{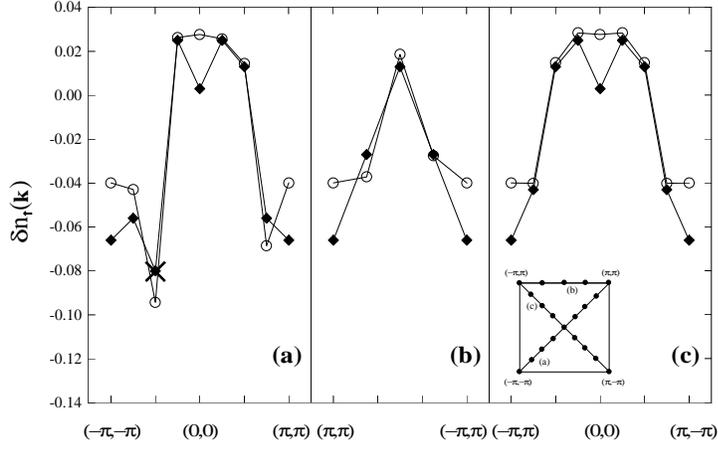}}}
\end{picture}
\caption{Comparison of the numerical (open circles) and analytical
(filled diamonds) results for 
$\langle\delta n_{{\bf k}\uparrow}\rangle$ in the
single-hole ground state at $J/t=0.3$ along the lines shown in the
inset ($(-\pi,-\pi)\rightarrow(\pi,\pi)
\rightarrow(-\pi,\pi)\rightarrow(\pi,-\pi)$). 
Solid lines are a guides to the eye.}
\label{n1upa}
\end{figure}

\begin{figure}
\unitlength1cm
\epsfxsize=7cm
\begin{picture}(4,9)
\put(3,1.5){\rotate[r]{\epsffile{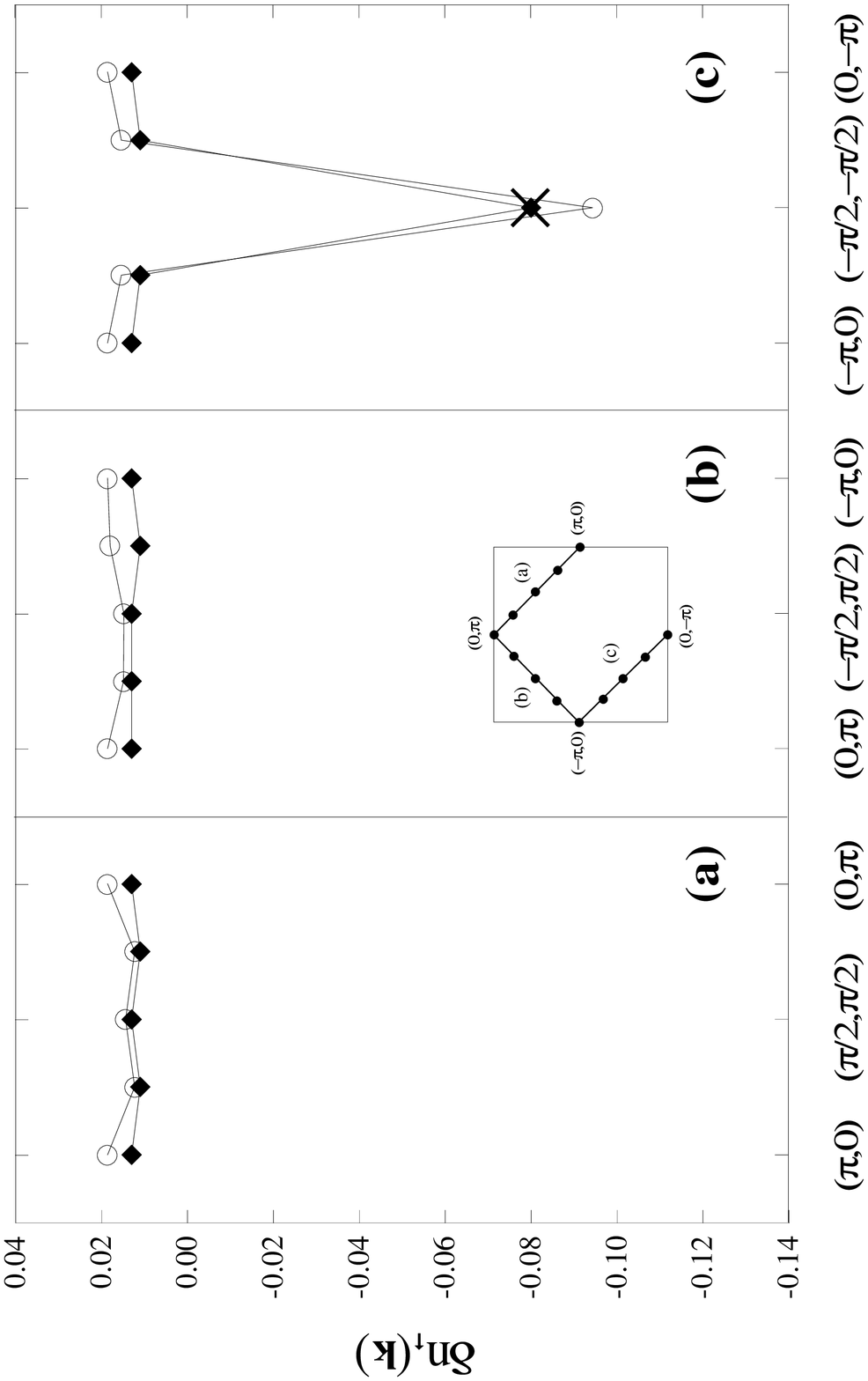}}}
\end{picture}
\caption{The same as in Fig.~\ref{n1upa} along the lines 
($(\pi,0)\rightarrow(0,\pi)\rightarrow(-\pi,0)\rightarrow(0,-\pi)$).}
\label{n1upb}
\end{figure}

\begin{figure}
\unitlength1cm
\epsfxsize=7cm
\begin{picture}(4,8)
\put(3,0.5){\rotate[r]{\epsffile{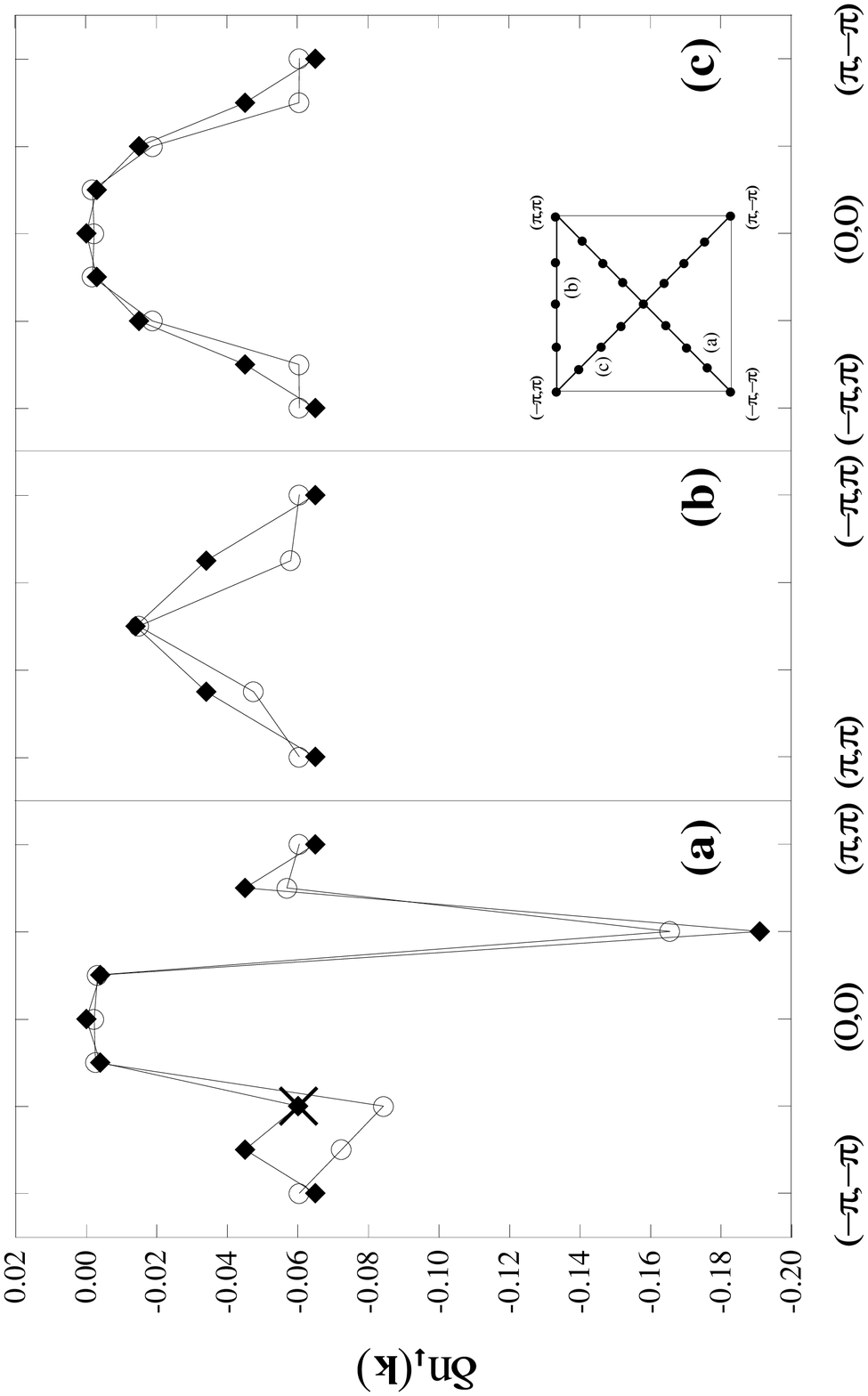}}}
\end{picture}
\caption{Comparison of the numerical (open circles) and analytical
(filled diamonds) results for 
$\langle\delta n_{{\bf k}\downarrow}\rangle$ in the
single-hole ground state at $J/t=0.3$ along the lines shown in the
inset ($(-\pi,-\pi)\rightarrow(\pi,\pi)
\rightarrow(-\pi,\pi)\rightarrow(\pi,-\pi)$). 
Solid lines are a guides to the eye.}
\label{n1downa}
\end{figure}

\begin{figure}
\unitlength1cm
\epsfxsize=7cm
\begin{picture}(4,8)
\put(3,0.5){\rotate[r]{\epsffile{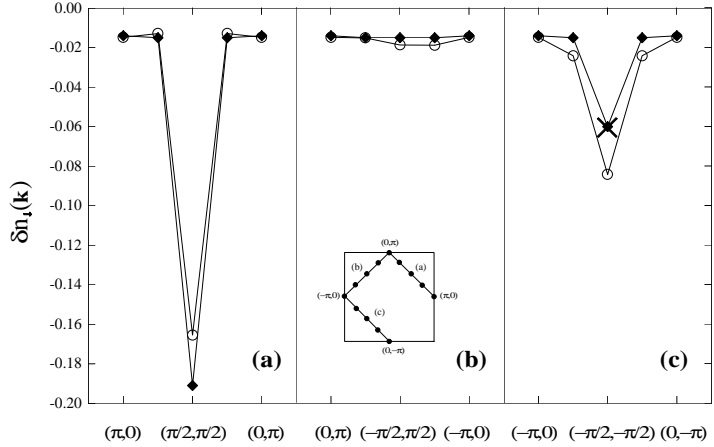}}}
\end{picture}
\caption{The same as in Fig.~\ref{n1downa} along the lines 
($(\pi,0)\rightarrow(0,\pi)\rightarrow(-\pi,0)\rightarrow(0,-\pi)$).}
\label{n1downb}
\end{figure}

Figures \ref{n1upa},\ref{n1upb},\ref{n1downa},\ref{n1downb} show our
analytical results for the single-hole EMDF (Eq. (\ref{4q})) together
with the 32-site ED data. Solid lines are guides to the eye. The
agreement is excellent for both spin directions. The differences between 
the $\langle \delta n_{{\bf k}\uparrow} \rangle$ 
numerical and analytical data
at $(0,0)$ and $(\pi,\pi)$ can be attributed to the fact that the CT
quasiparticle residue $Z_{\bf k}=a^2_{\bf k}$ at these points is
larger than the ``exact'' 
values ({\it e.g.}, SCBA). As one can see from 
Eq. (\ref{4q}), this leads to lower values of 
$\langle \delta n_{{\bf k}\uparrow} \rangle$. The agreement of the 
$\langle \delta n_{{\bf k}\downarrow} \rangle$ 
quantities away from
${\bf k} = {\bf P}$ is better because the role
of the background does not depend on $a_{\bf k}$. 
The ``antidips'' in the analytical results are marked by the cross
notifying that these points were calculated from Eq.~(\ref{4q}) using
finite gap value in the magnon spectrum \cite{gap}.

As we discussed in Sec. IV, the internal structure of the spin polaron
is made evident in the EMDF through the ``normal'' and ``interstring''
terms. The normal terms reflect the distribution of the hole
inside of the spin-polaron wave function, {\it viz.} strings
of different length. Interstring matrix elements are non-zero for 
$\langle\delta n_{\bf k} \rangle$ due to the specific structure of the spin
polaron. They make a contribution to the EMDF 
Eqs.(\ref{3q}),(\ref{4q}) which is asymmetric 
under the transformation ${\bf k}\rightarrow {\bf k}+{\bf Q}$. 
These asymmetric terms are responsible for the dome shape of 
$\langle\delta n_{{\bf k},\sigma}\rangle$, 
as was proposed in Ref. \cite{EB}, thus showing that it is not
related to a Fermi surface signature.

The $J/t$ dependence of the single-hole EMDF data has been extensively
studied in Ref. \cite{EdOh1} for smaller systems. As we already noted
in Sec. IV, Eq.~(\ref{4q}) naturally describes the results of these studies
and of the observations made in Sec. III.A.
To be specific, the depth of the 
``dip" is proportional to $Z_{\bf P}$, and thus
$\langle\delta n_{{\bf P},\downarrow}\rangle$ must follow the $J/t$
dependence of $Z_{\bf P}$ (result of Ref. \cite{EdOh1}). Also, the
background must be getting weaker for larger $J/t$ because less hole
weight is accumulated in the string cloud.
The smaller values and stronger 
$J/t$ dependence of $\Delta n_{\downarrow}$ and $\Delta
n_{anti,\downarrow}$ are due to 
the second and third components of the
spin-polaron wave function involved in the formation of the
$\langle\delta n_{{\bf k},\downarrow}\rangle$ background,
which are more sensitive to $J/t$. 

Now we consider two hole results, beginning with the binding energy.
This quantity for $d$-wave bound states was obtained numerically
and analytically at a variety of $J/t$. For $J/t = 0.3$ it was found that 
$E_b^{ED}=-0.05t$ and $E_b^{CT}=-0.02t$. As discussed before, the absence 
of a simple scaling law for $E_b$ does not allow one to produce a reliable
estimate of its thermodynamic limit at small
$J/t$. Nevertheless, we believe that the close agreement of the energies 
supports the idea that the systems under study represent the same physics.
A simple $1/N$ FSS for the larger $J/t=1.0$, where the size of the bound 
state is smaller and it is hoped that such a 
FSS will be more credible, gives a 
thermodynamic value of $E_b^{ED}\simeq -0.32t$ (Fig.\ref{En_2}(b)),
which is very close to the theoretical result $E_b^{CT}=-0.38t$.

Figure \ref{C_r}(a) shows our results for the hole-hole correlation function 
$C(r)$ for two holes in the $d$-wave bound state, for $J/t=0.3$. This 
expectation value is calculated firstly for a bulk lattice, and then mapped 
onto the equivalent sites of a 32-site cluster with periodic boundary
conditions. 
 This enforces that the 
analytical work approximates some of the finite-size effects of our 
ED numerics, and facilitates a more natural comparison between the two.
Very similar trends are found in both results, with the correlation function
decreasing quite similarly with the distance.

Both numerical and analytical results data show 
that about 45\% of the time the
holes prefer to stay at the nearest and next-nearest neighbour distances. 
That our analytical work produces such behaviour is not inconsistent with 
our statement regarding the form of $\Delta^{d}_{\bf p}$ Eq. (\ref{5a}): 
the {\it centres} of the polarons are indeed restricted to be on opposite 
sublattices, but the holes are almost equally distributed on both sublattices,
with the maximum probability of separation being at $\sqrt{2}$. In fact, our 
analysis of the harmonics in $\Delta^{d}_{\bf p}$ given in Table I shows
that about 40\% of the {\it polarons} in the bound state are separated by
one lattice constant. Thus, the peak of $C(r)$ at $r=\sqrt{2}$ arises
from the components of the wave function with
``strings" of length one. Clearly, the spin-polaron picture provides 
a natural explanation for the $\sqrt{2}$-paradox found here and in
earlier numerical studies. \\

Table I. {\small Amplitudes ($C_{2n-1,2m}$) and weights ($C^2_{2n-1,2m}$)
of the harmonics in the 
spin-polaron bound state wave function Eq. (\ref{5a}).
Weights are directly related to the polaron-polaron spatial
distribution function:
$P(r_{ll^{\prime}})=\frac{1}{z_{ll^{\prime}}}C^2_{ll^{\prime}}$, 
$z_{ll^{\prime}}=4;8$ is the
coordination number. $l$ and$l^{\prime}$ belong to different
sublattices}\\ \\
\begin{tabular}{ccc}\hline
\ \hskip 1.cm$(2n-1,2m)$ \hskip 1.cm \  &
\ \hskip 0.7cm $C_{2n-1,2m}$ \hskip 0.7cm \
&\ \hskip 0.7cm  $C^2_{2n-1,2m}$, \%   \hskip 0.7cm \  \\ 
\hline 
$(1,0)$       & 0.642  & 41.3 \\ 
$(1,2)+(1,-2)$ & 0.215  & 4.6  \\ 
$(3,0)$       & 0.444  & 19.7  \\ 
$(3,2)+(3,-2)$ & 0.320  & 10.2  \\ 
$(5,0)$       & 0.240  & 5.8  \\ 
\hline
\end{tabular}
\vskip 0.5cm
 
We believe that there are two reasons for the analytical $C(1)$ being 
slightly larger than $C(\sqrt{2})$. A treatment of the $t$-$J$ model based on 
the spinless hole representation involves some unphysical states with the 
hole and spin excitations being at the same site. 
The number of processes leading to such states increases when the
polarons are close to each other and hence $C(1)$ grows. Secondly,
the CT approach slightly overestimates bare hole weight.

An additional maximum in our analytical $C(r)$ at $r=3$ is closely connected to
the second important harmonic in the CT $d_{x^2-y^2}$ bound state $\sim
(\cos(3p_x)-\cos(3p_y))$. We cannot explain the absence of such a peak 
in the ED results --- for example, we have been unable to estimate the
finite size effect on individual harmonics in the two-hole wave function.

Figure \ref{C_r}(b) demonstrates a better agreement at $J/t=0.8$ when
the size of the bound state is small. In
this case the correlation fall rapidly with
distance and thus the ``bulk to cluster'' mapping does not alter the
analytical data. Therefore, in the large $J/t$ limit we find the
expected result that a spin-polaron approach adequately describes
the physics.

\begin{figure}
\unitlength1cm
\epsfxsize=7cm
\begin{picture}(8,9)
\put(3,1.5){\rotate[r]{\epsffile{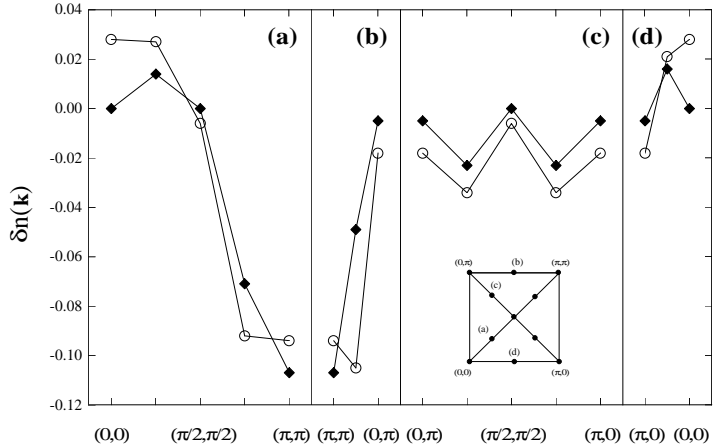}}}
\end{picture}
\caption{Comparison of the numerical (open circles) and analytical
(filled diamonds) results for 
$\langle\delta n_{{\bf k}\downarrow}\rangle$ in the
two-hole ground state at $J/t=0.3$ along the lines shown in the
inset ($(0,0)\rightarrow(\pi,\pi)\rightarrow(0,\pi)
\rightarrow(\pi,0)\rightarrow(0,0)$).  
Solid lines are a guides to the eye.}
\label{nq2comp}
\end{figure}

The EMDF for two holes shows an equally satisfactory comparison, as seen in
Figure \ref{nq2comp}.
The behaviour of $\langle\delta 
n_{\bf k} \rangle$ involves the combined effects of
(i) the internal structure of the polarons, (ii) the $d$-wave symmetry bound 
state,  and (iii) higher harmonics in the bound-state wave function.
We now elaborate on these features.

As we argued in Sec. III.B, the quantity $\Delta n^{2hole}= (\langle
n_{(0,0)}\rangle -\langle n_{(\pi,\pi)}\rangle)\simeq 
(\Delta n^{1hole}_{\uparrow}+\Delta n^{1hole}_{\downarrow})$ 
shows that the overall background deviation is mainly irrelevant to the
bound-state ${\bf k}$-structure. The worst agreement between ED and CT
analytics for the $(0,0)$ and $(\pi,\pi)$ points (Fig. \ref{nq2comp})
is again due to the $Z_{{\bf k}=0}$ problem within the CT approach.

Next we focus on the features along the ABZ boundary which, as we
discussed before, are not disguised by kinematic effects
(all asymmetric and most of the ``normal'' terms are zero on this
line) and can
be directly related to the form of $\Delta^{d}_{\bf p}$ in Eq. (\ref{5a}). 
Our analytical and numerical $\langle\delta n_{\bf k} \rangle$ results have 
a local maximum at ${\bf k} = (\pi/2,\pi/2)$, and have minima at 
$(3\pi/4,\pi/4)$ and $(\pi/4,3\pi/4)$. The first feature can be explained 
by the $d$-wave symmetry of the bound state. The EMDF is reduced from its
half-filled value of 1/2 when holes occupy that momentum state. 
However, as shown in Eq. (\ref{3w}), the EMDF 
consists of terms proportional to $(\Delta^{d}_{\bf k})^2$, 
which is identically zero at $(\pi/2,\pi/2)$. Thus,
$\langle\delta n_{\bf k} \rangle$ 
must show a local maximum (=0) at this wave vector, 
so one cannot find any direct
remnant of hole pockets for $d_{x^2-y^2}$-wave symmetry bound state.

A second feature that we observe in both analytical and numerical
results, {\em viz.} the minimum along the 
ABZ boundary between $(\pi,0)$ and $(\pi/2,\pi/2)$, can be related
to the particular form of $\Delta^{d}_{\bf p}$. Analytically, this 
quantity has large and apparently important higher harmonics (see
Table I and Eq. (\ref{5a})), and it is the competition between the
different harmonics that produces the maximum hole number between
$(\pi/2,\pi/2)$ and $(3\pi/4,\pi/4)$. 
Our analytical work shows that the hole number is actually maximized
along the ABZ boundary very close to $(\pi/2,\pi/2)$, roughly 
at $(0.45 \pi, 0.55 \pi)$. It is unclear if experiments could resolve 
this feature.

\section{Conclusions}

Summarizing, we have presented new ED numerical data for up to two holes
in the $t$-$J$ model for the largest cluster for which such calculations 
can be completed presently. 
Then, we compared these results with new analytical 
expressions based on the canonical transformation approach to the
$t$-$J$ model. We find
good agreement for the binding energy, the EMDF for one and two
holes, and the hole-hole spatial correlation function. We consider this
to lend strong support to the validity of the quasiparticle Hamiltonian
derived in Ref. \cite{bel97}, thus supporting the contention
that the spin-polaron description of the quasiparticles in the
$t$-$J$ model is correct at least at low hole concentration.

Certain characteristics in the correlation functions we studied
are direct consequences of some
features of the corresponding ground state wave functions.
For example, the dip in the single-hole spin 
$\downarrow$ EMDF is related to the centre of the spin polaron, whereas 
the dome structure of the 1- and 2-hole EMDF is due to the 
inter-string matrix elements of $\langle n_{{\bf k}\sigma} \rangle$. 
The large correlation 
between the holes in the $d$-wave bound state at a distance of $\sqrt{2}$ 
is due to the significant weight of the shortest string in the spin-polaron 
wave function. Analytical results for the FSS of the EMDF show a $1/N$ scaling 
at almost all ${\bf k}$-points except at the single-hole ground state {\bf P}
(Eqs. (\ref{3q}),(\ref{3w}),(\ref{4q})), in agreement with what is
shown in Sec. III for the ED data. Those ${\bf k}$-points which are
influenced by the long-range physics of the system are shown to
have more complicated scaling laws --- see Eqs. (\ref{nnq}),(\ref{nnq1}).
The role of the higher harmonics and the effect of the size of the bound
state on the EMDF and the hole-hole correlation function
for the two-hole problem are also discussed.

\begin{center}
{\bf ACKNOWLEDGMENTS}
\end{center}

We would like to thank Martin Letz, Frank Marsiglio, T.-K. Ng and Oleg Sushkov
for helpful comments. We are grateful to R. Eder for 
correspondence and for supplying us with useful references.
This work was supported by the RGC
of Hong Kong, and the NSERC of Canada. The 32-site ED 
work was completed on the Intel Paragon at HKUST.

\newpage
\appendix{}
\section{EMDF and $C(r)$ for two-hole case}

Amplitudes of the components of the two-hole wave function (\ref{6a}) 
$A^{(n)}, B^{(n)}, C^{(n)}$ can be expressed through the $a,b,c$
components of the single-hole wave function (\ref{3a}):
\begin{eqnarray}
\label{A3}
&&A_{\bf k}^{(1)}=a_{\bf k}^2 \ , \ \ \ A_{\bf k,q}^{(2)}= 2 
b_{\bf k,q}b_{\bf -k+q,q} \ , \ \ \ A_{\bf k,q,q^{\prime}}^{(3)}
=c_{\bf k,q,q^{\prime}}c_{\bf -k+q+q^{\prime},q^{\prime},q} 
\nonumber\\
&&B_{\bf k,q}^{(1)}= -a_{\bf k} b_{\bf -k,q}\ , \ \ \ 
B_{\bf k,q,q^{\prime}}^{(2)}=b_{\bf k,q^{\prime}}b_{\bf
k-q^{\prime},q}b_{\bf -k+q^{\prime},q^{\prime}}\\
&&C_{\bf k,q,q^{\prime}}^{(1)}=\frac{1}{2}a_{\bf k} c_{\bf
k,q,q^{\prime}} \ , \ \ \ 
C_{\bf k,q,q^{\prime}}^{(2)}= b_{\bf k,q}b_{\bf -k,q^{\prime}}  .
\nonumber 
\end{eqnarray}
$a_{\bf k}, b_{\bf k,q}$, and $c_{\bf k,q,q^{\prime}}$ within the CT
approach are the products of $M_{\bf k,q}$'s (\ref{1a}), $f_{\bf k}$'s, and
different integrals of their combinations as described in
Ref. \cite{bel97}. $f_{\bf k}$ is the transformation parameter
obtained from Eq. (\ref{N7}).

Using Eq. (\ref{4s}) and the bound state wave function (\ref{6a})
one can get EMDF for the two-hole ground state
\begin{eqnarray}
\label{A1}
&&\langle n_{\bf k}\rangle \simeq \frac{1}{2}+\frac{1}{N}\biggl[
-\langle\delta n^{even}_{\bf k}\rangle
+2\langle\delta n^{odd}_{\bf k}\rangle\biggr]
\nonumber\\
&&\phantom{\langle n_{\bf k}\simeq\rangle}
\langle\delta n^{even}_{\bf k}\rangle=FA^2_{\bf k} + \sum_{\bf q}FB^2_{\bf k,q}
+ \sum_{\bf q, q^{\prime}}FC^2_{\bf k,q,q^{\prime}} \\
&&\phantom{\langle n_{\bf k}\simeq\rangle}
\langle\delta n^{odd}_{\bf k}\rangle
=\sum_{\bf q} FB_{\bf k,q} U_{\bf q} FA_{\bf k+q} -
FA_{\bf k} \sum_{\bf q}V_{\bf q}FB_{\bf k,q}\nonumber\\
&&\phantom{\langle n_{\bf k}\simeq\rangle}
\phantom{\langle\delta n^{odd}_{\bf k}\rangle=}
-\sum_{\bf q, q^{\prime}}FC_{\bf k,q,q^{\prime}}U_{\bf q^{\prime}} 
FB_{\bf k+q+q^{\prime},-q} + \sum_{\bf q^{\prime}}FB_{\bf k,q^{\prime}}
\sum_{\bf q}V_{\bf q} FC_{\bf k+q^{\prime},-q,-q^{\prime}}\nonumber ,
\end{eqnarray}
where
\begin{eqnarray}
\label{A2}
&&FA_{\bf k}=\Delta^{d}_{\bf k}A_{\bf k}^{(1)} +
\sum_{\bf q} \Delta^{d}_{\bf k+q} A_{\bf k+q,q}^{(2)}
+\sum_{\bf q,q^{\prime}} \Delta^{d}_{\bf k+q+q^{\prime}}
A_{\bf k+q+q^{\prime},q,q^{\prime}}^{(3)}\nonumber\\
&&FB_{\bf k,q}=\Delta^{d}_{\bf k}B_{\bf k,q}^{(1)} -
\Delta^{d}_{\bf k+q}B_{\bf k+q,-q}^{(1)}+
\sum_{\bf q^{\prime}}\left( \Delta^{d}_{\bf k+q+q^{\prime}} 
B_{\bf k+q+q^{\prime},q,q^{\prime}}^{(2)}
-\Delta^{d}_{\bf k+q^{\prime}} B_{\bf k+q^{\prime},-q,q^{\prime}}^{(2)}
\right)\\
&&FC_{\bf k,q,q^{\prime}}=\Delta^{d}_{\bf k+q+q^{\prime}}
C_{\bf k+q+q^{\prime},q,q^{\prime}}^{(1)} +
\Delta^{d}_{\bf k}C_{\bf -k,q^{\prime},q}^{(1)}-
\Delta^{d}_{\bf k+q^{\prime}} C_{\bf k+q^{\prime},-q,-q^{\prime}}^{(2)} .
\nonumber 
\end{eqnarray}
Hole-hole correlation function (\ref{4s}) in the bound state
(\ref{6a}) is given by
\begin{eqnarray}
\label{A4}
&&C(r_{ij})\simeq \cases{
C^{00}(r_{ij})+C^{22}(r_{ij})=\langle 2|n^f_i n^g_j|2\rangle ,
& when $i+j=2n-1$ \cr
C^{11}(r_{ij})=\langle 2|n^f_i n^f_j|2\rangle, & when $i+j=2n$\cr}
\nonumber \\
&&C^{00}(r_{ij})=\biggl(\sum_{\bf k}\Delta^{d}_{\bf k}
\biggl\{ A_{\bf k}^{(1)}\cos[{\bf kr}_{ij}] + \sum_{\bf q}A_{\bf
k,q}^{(2)}\cos[({\bf k-q}){\bf r}_{ij}] 
+ \sum_{\bf q, q^{\prime}}A_{\bf k,q,q^{\prime}}^{(3)}\cos[({\bf
k-q-q^{\prime}}){\bf r}_{ij}]\biggr\}\biggr)^2 \nonumber \\
&&C^{11}(r_{ij})=\sum_{\bf k,p}\Delta^{d}_{\bf k}\Delta^{d}_{\bf p}
\sum_{\bf q}\biggl\{ B_{\bf k,q}^{(1)}B_{\bf p,q}^{(1)}\bigl(\cos[({\bf
k-p}){\bf r}_{ij}]-\cos[({\bf k+p+q}){\bf r}_{ij}]\bigr) \\
&&\phantom{C^{11}(r_{ij})=\sum_{\bf k,p}\Delta^{d}_{\bf k}
\Delta^{d}_{\bf p}\sum_{\bf q}\biggl\{ }
+ 2 B_{\bf k,q}^{(1)} \sum_{\bf q^{\prime}}B_{\bf k,q,q^{\prime}}^{(2)}
\bigl(\cos[({\bf k-p+q+q^{\prime}}){\bf r}_{ij}]-\cos[({\bf
k+p-q^{\prime}}){\bf r}_{ij}]\bigr)\biggr\}
\nonumber \\
&&C^{22}(r_{ij})=\sum_{\bf k,p}\Delta^{d}_{\bf k}\Delta^{d}_{\bf p}
\sum_{\bf q,q^{\prime}}\biggl\{\biggl( C_{\bf k,q,q^{\prime}}^{(1)}
C_{\bf p,q,q^{\prime}}^{(1)}+C_{\bf -k,q^{\prime},q}^{(1)}
C_{\bf -p,q^{\prime},q}^{(1)}+C_{\bf k,q,q^{\prime}}^{(2)}
C_{\bf p,q,q^{\prime}}^{(2)}\biggr)\cos[({\bf k-p}){\bf r}_{ij}]
\nonumber \\
&&\phantom{C^{22}(r_{ij})=\sum_{\bf k,p}\Delta^{d}_{\bf k}\Delta^{d}_{\bf p}
\sum_{\bf q,q^{\prime}}\biggl\{ }
+ 2 C_{\bf k,q^{\prime},q}^{(1)} C_{\bf -p,q^{\prime},q}^{(1)} 
\cos[({\bf k-p+q+q^{\prime}}){\bf r}_{ij}]-
2 C_{\bf k,q^{\prime},q}^{(2)} C_{\bf p,q,q^{\prime}}^{(1)}
\cos[({\bf k+p-q^{\prime}}){\bf r}_{ij}]
\nonumber \\
&&\phantom{C^{22}(r_{ij})=\sum_{\bf k,p}\Delta^{d}_{\bf k}\Delta^{d}_{\bf p}
\sum_{\bf q,q^{\prime}}\biggl\{ }
-2 C_{\bf k,q,q^{\prime}}^{(2)} C_{\bf -p,q^{\prime},q}^{(1)}
\cos[({\bf k+p+q^{\prime}}){\bf r}_{ij}] \biggr\}
\nonumber 
\end{eqnarray}

\newpage


%


\begin{references}
\bibitem{gap} The value of the gap can be calculated from
Eq.~(\ref{E_N}). The result in leading order in $1/N$ is 
$\Delta E = 0.498J$. Taking into account the $1/\sqrt{N}$ correction
change this result to $\Delta E = 0.237J$. In the calculations we used
the ED value of the gap, $\Delta E^{ED} = 0.315J$. We believe that the
use of this latter relation
provides a reliable estimation of the EMDF quantities.
\end{references}
\end{document}